\documentclass{sn-jnl}
\usepackage[labelfont=bf, font=small]{caption}
\usepackage{lmodern}
\usepackage{graphicx}%
\usepackage{multirow}%
\usepackage{amsmath,amssymb,amsfonts}%
\usepackage{amsthm}%
\usepackage{mathrsfs}%
\usepackage[title]{appendix}%
\usepackage{xcolor}%
\usepackage{textcomp}%
\usepackage{colortbl}
\usepackage{manyfoot}%
\usepackage{booktabs}%
\usepackage{algorithm}%
\usepackage{algorithmicx}%
\usepackage{algpseudocode}%
\usepackage{listings}%

\usepackage{comment}
\usepackage{tabularx}
\usepackage{array}
\usepackage{float}
\usepackage{booktabs}
\usepackage{adjustbox}

\usepackage{caption}
\usepackage{subcaption}
\usepackage{hyperref}
\usepackage{enumitem}

\begin{document}

\title[Assembly Theory Reduced to Shannon Entropy]{\vspace{-3.3cm}\Large Assembly Theory Reduced to Shannon Entropy and Rendered Redundant by Naive Statistical Algorithms\vspace{-0.5cm}}


\author[4]{\fnm{\textnormal{} Luan} \sur{\textnormal Ozelim}}

\author[4,5,6]{\fnm{\textnormal Abicumaran} \sur{\textnormal Uthamacumaran}}

\author[1,2]{\fnm{\textnormal Felipe S.} \sur{\textnormal Abrah\~{a}o}}

\author[4]{\fnm{\textnormal Santiago} \sur{\textnormal Hern\'andez-Orozco}}

\author[3,4]{\fnm{\textnormal Narsis A.} \sur{\textnormal Kiani}}

\author[7]{\fnm{\textnormal Jesper} \sur{\textnormal Tegn\'er}}

\author[  4,8,9]{\fnm{\textnormal Hector} \sur{\textnormal Zenil}\footnote{Corresponding author: hector.zenil@kcl.ac.uk}}

\affil[1]{\small Center for Logic, Epistemology and the History of Science, University of Campinas, Brazil\small}
\affil[2]{\small DEXL, National Laboratory for Scientific Computing, Brazil\small}
\affil[3]{\small Department of Oncology-Pathology, Center for Molecular Medicine, Karolinska Institutet, Sweden\small}
\affil[4]{\small Algorithmic Dynamics Lab, Karolinska Institutet \& King's College London, Sweden \& U.K\small}
\affil[5]{\small Department of Physics and Psychology, Concordia University, Canada\small}
\affil[6]{\small McGill University, Neurosurgical Simulation and AI Learning Centre, Canada\small}
\affil[7]{\small Living Systems Lab, King Abdullah University of Science and Technology, KSA\small}
\affil[8]{\small Algorithmic Dynamics Lab, Department of Biomedical Computing, School of Biomedical Engineering and Imaging Sciences \& King's Institute for AI, King's College London, U.K\small}
\affil[9]{\small Cancer Research Group, The Francis Crick Institute, London, U.K\vspace{-1.5cm}}

%
%


\abstract{Assembly Theory (AT) and its central measure, the assembly index (Ai), represent an invaluable opportunity to address some of the most persistent and widespread conflations and misconceptions about computability and complexity theory in science. The AT defence embodies several common concurrent misconceptions that pile on each other: the belief that Turing machines impose artefactual constraints, the mischaracterisation of Kolmogorov complexity as inapplicable, and the claims around Ai as different from Shannon entropy or compression algorithms. Here we show that the new arguments advanced by the AT group in their defence, are based on misleading and incomplete experiments that, when completed, show the extent of the correlations and overlapping with popular statistical compression algorithms, conforming with the mathematical equivalence to Shannon entropy previously mathematically proved and reported, which remains undisputed. Through theoretical and empirical analysis, we show that Ai does not offer a path towards fundamental novel causal or informational insights beyond what existing statistical frameworks already offer. Rather than offering a unifying theory of life as the AT authors suggest, we argue that AT obfuscates the field and provides a cautionary example of how the accumulation of conceptual mistakes can lead to a misleading theory. Finally, we show that Ai is a particular limited case of another complexity metric based on algorithmic (Kolmogorov) complexity, consisting of decomposing an object into its causal blocks that goes beyond, and outperforms, AT.\\

}

\maketitle


\section{What a ZIP File Cannot Tell About Life}
Assembly Theory (AT) is a framework introduced recently that seeks to explain and quantify selection and evolution, and, consequently, life, by focussing on the minimal number of steps required to build an object of interest from its basic components. According to its authors, it emphasises historical and causal construction pathways rather than static descriptions of structure, proposing that objects with high ``assembly indices'' carry evidence of selection processes and cumulative history~\cite{Marshall2021,Sharma2023}. By quantifying how hard it is to recreate something from simpler parts, AT claims to provide a metric for identifying signatures of life and distinguishing naturally occurring complexity from randomness. Although these descriptions sound exactly how algorithmic (Kolmogorov) complexity, Shannon entropy, and popular statistical compression approaches have tried to characterise complexity and life before, the authors insist that theirs is a different and novel approach.

According to the AT authors, the number of exact copies in an object is a measure of its selective content~\cite{Sharma2023} and that this number is key to characterising life on Earth and beyond~\cite{Marshall2021}, explaining time, matter and more recently even the expansion of the universe and cosmic inflation, according to the authors~\cite{templeton}.

The primary goal of this paper is to clarify some aspects of AT and their Ai under new defence arguments as published in~\cite{kempes2024}. In particular, its equivalence to compression algorithms and Shannon entropy. In doing so, we will take advantage to address common misconceptions of computability and complexity theories in science, of which AT is a quintessential example and an opportunity as a case study.

In a recent article~\cite{kempes2024}, responding to some of these criticisms, putative evidence was offered to claim that AT and its Ai metric are not statistical compression algorithms or at least not popular algorithms or equivalent to Shannon entropy.

Early published findings demonstrated that Assembly Theory (AT) and a version of their Assembly Index (Ai) called molecular assembly or MA when applied to molecular complexity, performed similarly or suboptimally compared to simple statistical compression algorithms such as LZW in their own goal at separating organic from non-organic molecules, with a strong Pearson correlation of 0.874~\cite{Uthamacumaran2024}. In what follows, Ai and MA will be used interchangeably, since Ai includes MA, and MA is a specific case of Ai for molecular complexity.

In the same paper~\cite{Uthamacumaran2024}, it was also shown that LZW and other compression methods, including those based on Shannon entropy ($\mathbf{H}$), were able to distinguish the same organic from inorganic compounds using the AT authors' own mass spectral data as effectively as, or better than, the assembly index. We concluded that Ai~\cite{Uthamacumaran2024}, and therefore AT, offered no substantial improvement over existing statistical methods for identifying biosignatures or separating living from nonliving systems, which has been a central basis of the Assembly Theory authors' claim that their theory explained evolution and selection, and unified biology and physics~\cite{Sharma2023}. However, the authors had never offered such basic control experiments in any of their previous papers, until now~\cite{kempes2024}. However, we will argue that the authors only offered a handful of examples that are misleading and incomplete, and they did not address the mathematical proof and the overwhelming evidence that their results can be all replicated by simple existing algorithms, including LZW and Shannon entropy and that there is no justification for them to believe that nature compresses objects as they propose than in any other form, but we do also have evidence that nature does not operate like AT given that AT and Ai cannot account for basic chemical and biological operations.

Here we also prove that the Block Decomposition Method (BDM) that we introduced in the early 2010s~\cite{zenilbdm}, which does not shy away from combining $\mathbf{H}$ and causal local structure by algorithmic probability~\cite{zenilctm,nmi2019}, properly subsumes statistical compression methods such as AT and its assembly index (denoted by Ai). BDM starts by counting identical copies but also other causal operations such as reversion, complementation, inversion, and other linear and nonlinear transformations---known to be widely used by evolution and selection (e.g. right-left symmetry)---that AT does not consider, among other causal (and nonstatistical) transformations that BDM may capture~\cite{nmi2019} but AT cannot.

\subsection{Preliminary Concepts}

Introduced by Lempel, Ziv, and Welch in the late 1970s, the LZW algorithm that carries their initials was the first dictionary-based algorithm after Huffman's first compression algorithm that started off the field of data compression~\cite{lzw}. Over time, many other algorithms, such as LZ77 and LZ78, were introduced. which in turn underlie popular compression algorithms such as ZIP, GZIP and PNG. All these algorithms are based on the same counting principle that AT and their Ai seem to suggest they have introduced. The differences between all those compression algorithms are small variations introduced for reasons such as efficiency that were not originally considered, and of which AT with their Ai is one.

A popular application of lossless compression algorithms like LZW and others in the LZ family is to estimate the algorithmic complexity (Kolmogorov or Kolmogorov-Chaitin), the accepted mathematical definition of randomness based on the length of the shortest computer programs capable of mechanistically generating an object. Formally, $\mathbf{K_U}(x)= \min\{|p|,U(p)=x\}$.

This metric is deeply connected to another fundamental concept in complexity theory, that of Algorithmic Probability (AP). As introduced by Solomonoff~\cite{solomonoff}, AP assigns a higher likelihood of mechanistic production to objects that can be produced by shorter programs and a lower probability to objects that do not have shorter descriptions. The Coding Theorem Method (CTM) connects these two metrics using the relation
$\mathbf{K}(x) \approx -\log_2 AP(x)$ known as Coding Theorem~\cite{zenilctm} where $AP(x)$ denotes the algorithmic probability of $x$, defined as
$AP(x) = \sum_{p : U(p) = x} 2^{-|p|}$ where $|p|$ is the length of the programme $p$ and $U$ an optimal universal reference constructor $U$ with the length of $U$ absorbed into a constant common to all measurements~\cite{Li1997}. The instantiation of CTM involves a (Levin) search that finds multiple (short) computer programs (or pathways) to generate $x$.

A method that extends CTM to approximate $\mathbf{K}$ via $AP$ called the Block Decomposition Method (BDM) divides larger objects into sub-blocks, estimating the complexity of each by counting the number of random programs able to reconstruct (assemble) the smaller objects, introducing a logarithmic penalty for repeated structures that effectively counts each block repetition. The result is a hybrid measure that in the `worse case', approximates Shannon entropy rate in linear time by simply counting assembly blocks and compressing the object optimally according to the rules of classical information theory or, in the best case, the same method finds small generative computer programs that provide a clue on the causal mechanisms and causal content that assembles such an object, thus combining the best of both classical and causal algorithmic information theories~\cite{zenilbdm}.

In its standard formulation, BDM decomposes an object $X$ into sub-blocks $x_i$ as follows:

\begin{equation}
\text{BDM}(X) = \sum_i CTM(x_i) + \log_2(n_i)
\end{equation}

where CTM stands for Coding Theorem Method, the search in software space of generative mechanisms explaining and generating each $x_i$ for each $i$ composing and re-assembling $X$~\cite{zenilctm}.

The Block Decomposition Method (BDM), introduced in 2013~\cite{zenilbibm,zenila,zenilbdm}, is an estimator of algorithmic information redundancies $\mathbf{K}$ based on the principle of decomposing an object into parts for which algorithms complexity estimates are aggregated. It counts exact copies in data as its most basic test for nonrandomness, but unlike purely statistical measures, BDM penalises for trivial repetitions that are also found in simple stochastic processes, such as rock formations or crystals~\cite{zenilbdm}. Estimation for parts is obtained using the Coding Theorem Method (CTM)~\cite{zenilctm} based on Algorithmic Probability~\cite{solomonoff,chaitin} (AP) which considers all statistical and algorithmic regularities in the data. 

Given that CTM can only find small programs, BDM extends its power by stitching these programs together to offer a causal explanation of the object as a whole.

Then, according to the principles of classical information theory, BDM characterises the information content of the entire object by adding the estimated (local) complexity and the (global) Shannon entropy ($\mathbf{H}$) values as described in Equation~\eqref{eqBDM}.

In what follows, we will show that naive statistical methods are able to replicate the results reported by Assembly Theory as original. Moreover, we will show that Ai is subsumed in BDM and achieves better results across multiple domains of AT application including but not exclusive of a greater separation between organic and inorganic compounds when used on the same data as provided by the AT group, in particular what they call `physical' data from their mass spectral values of molecular compounds.

\section{New Defence Arguments and Results}\label{sectionZIP}

In ~\cite{kempes2024}, the authors of Assembly Theory (AT) argue that being worse at compressing objects makes Ai different, as the assembly index quantifies selection and assumes no other existing statistical compressor can, recycling the argument (also partially rehashed in~\cite{Wolpert2024StochasticProcessTuring}) that Ai can deal with stochastic physical, chemical or biological objects while compression can only deal with computers, deterministic systems, and bits and programs.

The counterexample offered in their paper~\cite{kempes2024} amounts to the argument that Assembly Theory has the potential to implement better compression than LZW or ZIP for their purposes because by design it does not provide any means to compute it efficiently.

We recognise the recent refinement of the claims made by the authors of Assembly Theory~\cite{kempes2024}, who have moved from broader, unfounded assertions~\cite{fridman2024} to the more modest position that AT is a minor variation of a compression algorithm that belongs to a different (slower) time complexity class. However, unfounded claims in connection to their unique capabilities, even to characterise selection and evolution, remain and continue to be made.

\subsection{AT arguments against compression based on artefacts}

Regarding efforts to provide evidence with an example indicating some divergence from LZW, here we will show that such an example not only does not represent the general case or a correctly performed experiment, but also does not directly address the criticisms in~\cite{Uthamacumaran2024,abrahao2024}, including the mathematical proof showing equivalence with Shannon entropy.

Another argument made by the authors in~\cite{kempes2024} relates to the suboptimality of the assembly index as a compression algorithm. The authors of Assembly Theory have mistakenly suggested that one of our criticisms is that their assembly index is not an optimal compression algorithm. What we actually took issue with was that it was not presented as a compression scheme, not that it was nonoptimal.

In such arguments, the authors seem to misuse the concept of optimality---which is always relative to a given referential basis or context of analysis--- in favour of partial counter-arguments. LZW itself is only optimal as a Shannon entropy ($\mathbf{H}$) estimator in the limit, not as an algorithmic complexity ($\mathbf{K}$) estimator.

No statistical compression algorithm can achieve the optimality of algorithmic complexity, so we never expected this optimality from their assembly index.

The second part of the misunderstanding on the part of the authors of AT is that our criticism amounted to a claim that AT had the intention (or was designed) to capture the idea of optimal compression.

Intentions should not play a role in scientific debate, as using arguments about intentions to evaluate the validity or soundness of a theory or metric~\cite{kempes2024} is not a scientific discussion. Claims about differences solely based on the academic field in which a theory is rooted or on where the original motivations of scientists originate~\cite{Jirasek2024Assemblytheory,Sara2024BrianKeating} constitute a fallacy of origin.


One of our main criticisms is that the authors of AT have not presented any evidence that AT's nonoptimal compression algorithm is better than any other, especially in the face of our previous results replicating and improving upon their results at tasks such as separating organic from inorganic compounds~\cite{Uthamacumaran2024} on which most of their results and arguments are based on or originated from.

The mathematical motivation and mechanism behind LZ algorithms are the same as that of the assembly index (Ai), to count the number of exact repetitions in an object. If Assembly Theory has some pedagogical value, it is for its authors to make the case in a more restrained manner, eschewing the authors' claims but it is not methodologically different from Shannon entropy and the mathematical description of LZ algorithms~\cite{abrahao2024} and therefore AT with its Ai can be viewed as a method that leads to ZIP file compression that can tell little to nothing about the selection or evolution of an object beyond what Shannon entropy or compressing such an object with ZIP or GZIP tools, could not already tell.

Given that we have proven that almost any other statistical approach, including those originally designed to count copies as the authors of AT set out to do, can replicate their results or even improve upon them~\cite{Uthamacumaran2024}, and in the absence of empirical evidence, Assembly Theory would appear empty and insubstantial.

Moreover, as we have previously discussed in~\cite{abrahao2024} (later corroborated in \cite{Raubitzek2024AutocatalyticSetsAssembly}), we know that characterising selection without invoking the relationship of an agent interaction as selection if a function of an agent's environment and circumstances, such as competition for resources, is impossible for intrinsic complexity measures, such as those based on the assembly index. Furthermore, it makes no technical sense, abiotically or otherwise, as the transition to favour a certain system's traits would still need to be explained. These conflations with selection and evolution as used in a speculative prebiotic and biological context are an unnecessary obfuscation that only misleads readers.

The claim made in the Assembly Theory paper~\cite{kempes2024} that ``the assembly index is mistaken for being what is called a `special case' of algorithmic complexity without validation of the claim'' shows a profound misunderstanding of the basics of information theory, both classical and algorithmic. While proponents of AT distinguish between the concepts of ``copy number'' and the ``assembly index,'' our analysis focuses on the latter, which is the operational core of the theory in its computational form. The notion of copy number contributes conceptually to AT but, when implemented algorithmically, results in the same behaviour as Shannon entropy algorithms and LZ-style compression schemes.

We have formally demonstrated step-by-step not only that the assembly index and the assembly number are special-case approximations of algorithmic complexity, formalised in the language of---and based on---the results of algorithmic information theory (AIT)~\cite{Downey2010}.

While proponents of AT argue that Ai is intractable and thus distinct from compression algorithms like LZW, we emphasize that computational complexity (e.g., resources such as time or algorithmic cost) is separate from representational capacity. A method's intractability does not confer additional explanatory power.

The claim that the assembly index is different from dictionary-based compression because there is no dictionary is unfounded. The very book-keeping that tracks the order and frequency of molecular substructures \emph{is} the dictionary: the molecular sequence provides the alphabet, and the frequency counts instantiate the symbol---frequency table required for compression. In this formulation, the dictionary is not external but intrinsic to the data itself. This feature is not novel. Turing already showed that the boundary between programme and input disappears in universal computation, where the substrate can serve simultaneously as machine and as data. Other Turing-equivalent formalisms such as Post tag systems and the $\lambda$-calculus operate precisely in this manner, with no explicit external dictionary but with the rules and data co-encoded in the substrate. Far from distinguishing Ai from compression, this places it squarely within the standard mechanics of information representation as understood across multiple models of computation.

\subsection{Empirical Replication and Correlation Analysis}

\subsubsection{Analysis on Synthetic Strings}

The results in this section accord with our mathematical proof demonstrating that the assembly index (and any variation based on their concept of `copy number' based on counting identical copies in data) approximates, and is equivalent to, Shannon entropy via an LZ compression scheme and is, therefore, indistinguishable from Shannon entropy and LZ algorithms (of which LZW is one). It follows that results previously reported by Assembly Theory are, therefore, given their relationship to and implementation of Shannon entropy via LZ compression, not unique.

\begin{figure}[htp]
\centering
\subfloat[Complete correlation plot for the specific best counter example provided by AT authors]{%
\includegraphics[clip,width=0.6\columnwidth]{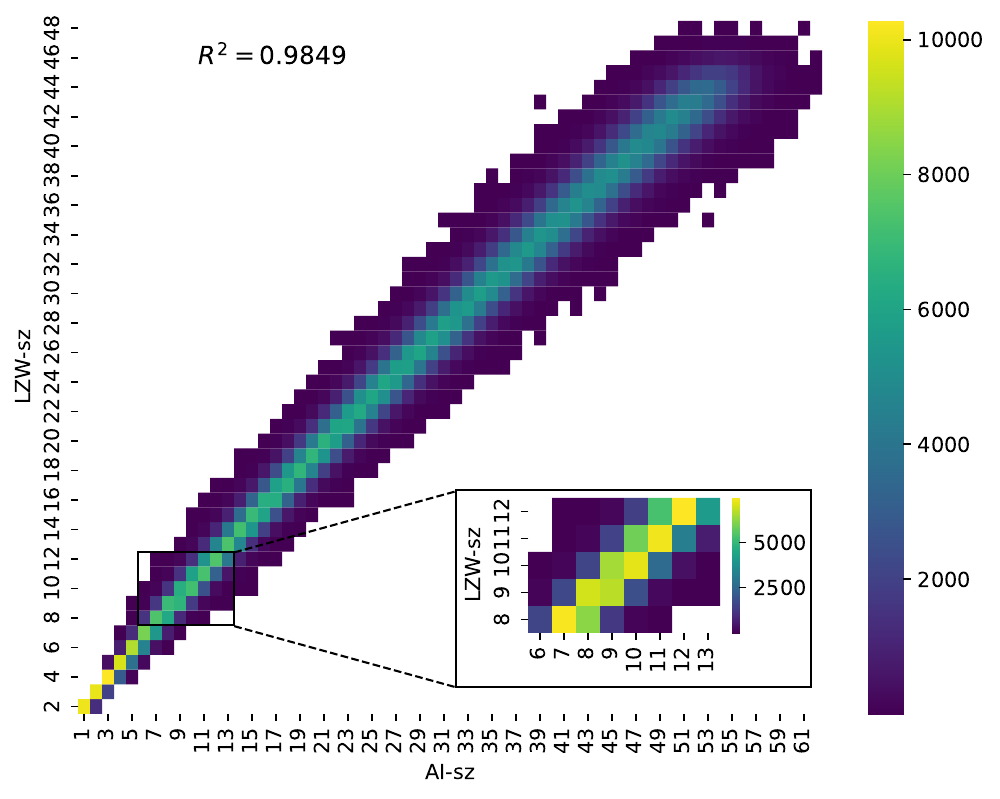}%
}\\
\subfloat[Parts that make up the inset image]{%
\includegraphics[clip,width=0.75\columnwidth]{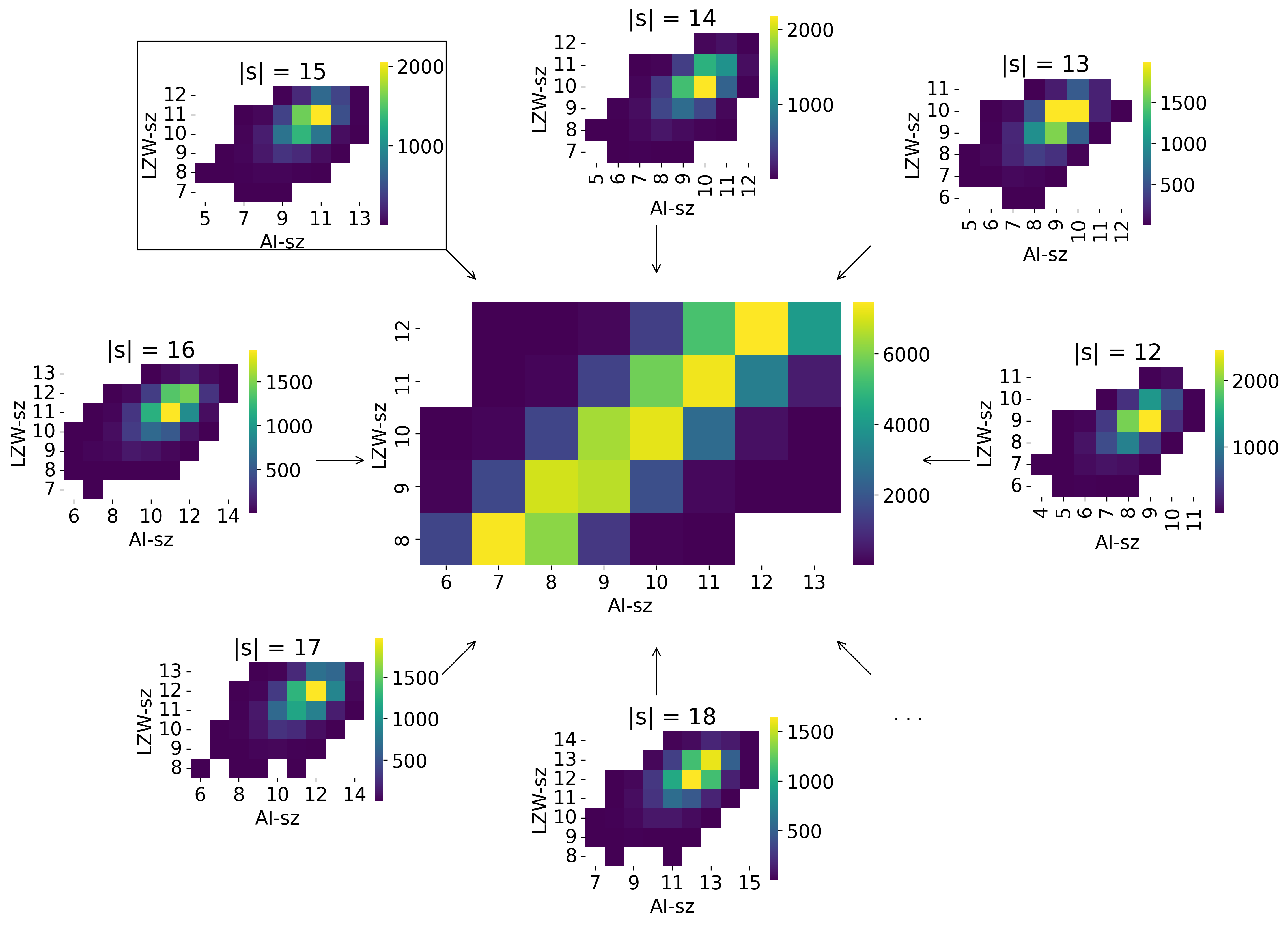}%
}
\caption{Main figure: correlation plot between the most popular statistical compression algorithm LZW (behind ZIP and PNG) and the assembly index (Ai) for a growing ZBC sequence (length size of 100 characters) and its random permutations following the experiment in~\cite{kempes2024}. The colour code corresponds to the number of strings which produced the pair (Ai,LZW) during the experiments. The authors of Assembly Theory only reported the component with string size equal to 15 (highlighted in (b)), in their effort to demonstrate the novelty of Ai compared to LZW~\cite{kempes2024} with a weak 0.25 correlation, without conducting or reporting on the full experiment. The apparent scattering of it is an artifact of the fixed scale. The quick asymptotic convergence to Spearman correlation 1 for multiple examples is provided in Table~\ref{tabpatterned} and its undisputed mathematical equivalence in~\cite{abrahao2024}. Unlike the claim in~\cite{kempes2024}, the assembly index correlation does not weaken with object size~\ref{tabpatterned}.}
\label{fig:heatmap}
\end{figure}

\begin{figure}[htbp]
\centering
\includegraphics[scale=0.23]{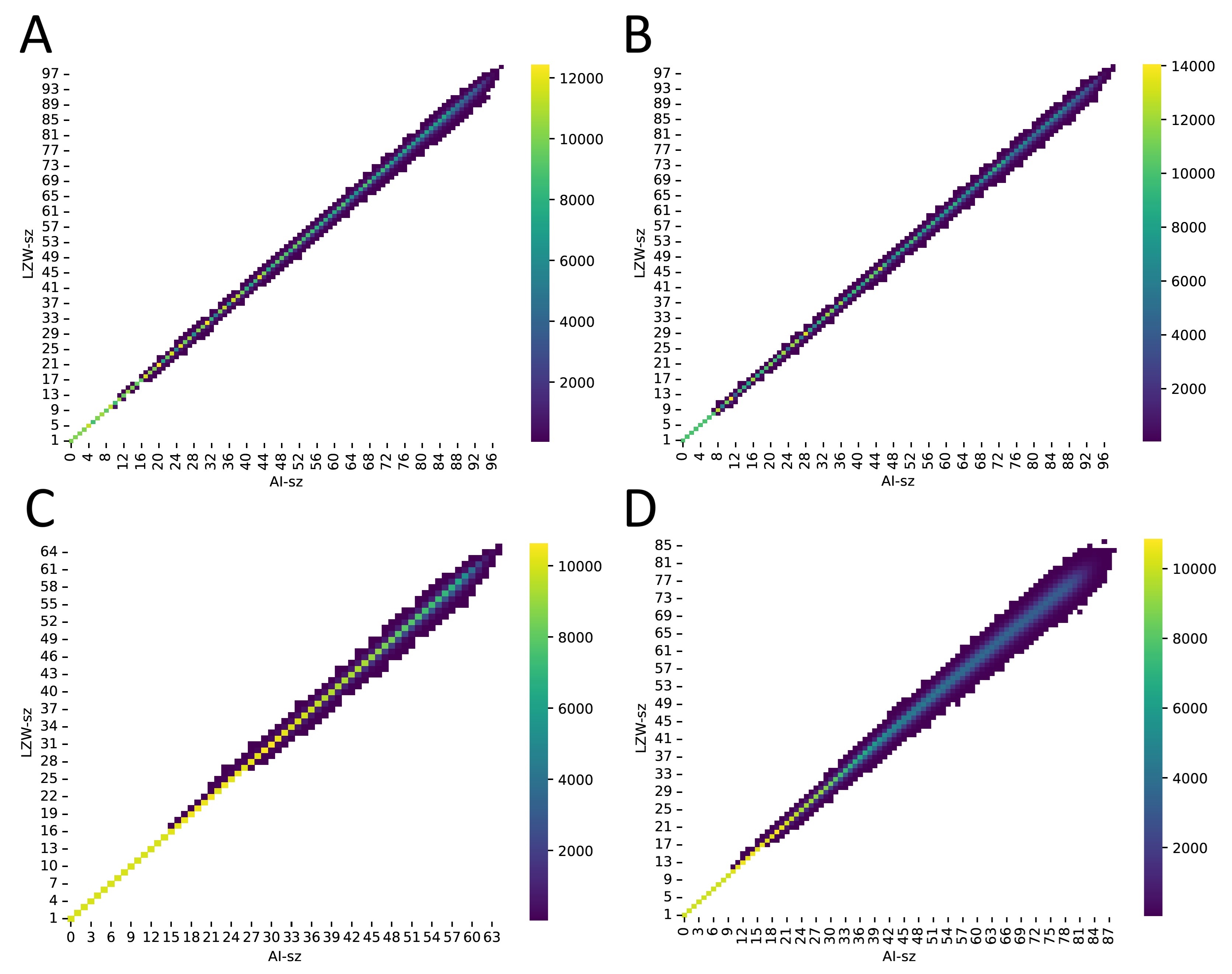}
\caption{Density correlation plots for strings of growing length confirm the convergent asymptotic behaviour between the assembly index and LZW. A and B are random sequences of length up to 100 while C and D are patterned sequences of repeated blocks of up to 100 characters. C shows a patterned block of ABCDE letters repeating, while D shows a ten-letter repeating block. The variance is lower in the random sequences, with growing size converging to very high correlations of 0.98-0.99, while in patterned sequences the correlations of the growing chain are high at about 0.9. Even when visually it may appear that the scattering increases for patterned strings, this is a visual artifact from the number of greater block lengths. As shown in the numerical Table~\ref{tabpatterned}, the correlation in all cases converges quick to 1 as a function of length as shown in Figs.~\ref{fig:sub12},\ref{fig:sub22}. But the exercise was futile given that we had already proven the mathematical equivalence in~\cite{abrahao2024}. Unlike the claim in~\cite{kempes2024}, the assembly index correlation does not weaken with object size as shown in Table~\ref{tabpatterned}.}
\label{main}
\end{figure}

The authors of AT offer a sequence of very simple examples as evidence in their favour. The first consists of the letters Z, B, and C arranged in long repetitive or random strings. Their main argument~\cite{kempes2024} is tested in Figs.~\ref{fig:heatmap} and~\ref{main}. The contrast is evident between the range of values that Ai---and indeed LZW and Shannon entropy---produces for the fixed-length permutations of the ZBC base string, and that produced by BDM, a measure constructed on fundamentally different or extensive principles, as opposed to simply counting number of repetitions as Ai, LZW and Shannon entropy do---a fact which explains their similarities.

The results for the growing sequences of the ZBC type, the patterned type and random strings are presented in Fig.~\ref{fig:heatmap}, showing the degree of lack of care and misleading selectivity in the results presented in~\cite{kempes2024}, with only partial results being reported, thus concealing the degree of alignment between LZW and Ai, as was reported and proved before~\cite{abrahao2024,Uthamacumaran2024}.

The analysis of Fig.~\ref{fig:heatmap} indicates that when the experiment is carried out only slightly further, the full picture and extent of the asymptotic convergence between Ai and LZW is revealed. This is in complete agreement with the theoretical expectation of the mathematical proof we offered before establishing the equivalence of Ai with Shannon entropy~\cite{abrahao2024} (via LZ). Effectively, this means that any previous or future positive (or negative) results obtained using AT would be due to its equivalence to statistical compression and Shannon entropy and not because of any particularity or advantage of AT or Ai. Due to the heuristics introduced in the calculation of Ai by the authors of AT, some deviations from the diagonal can be observed with the vanishing number of outliers as a function of length. Those variations do not prove the difference beyond local anomalies and do not grant AT or Ai any of the features the authors have represented it with, including claims related to the abilities of explaining selection and evolution.

\begin{figure}[htbp]
\centering
\begin{subfigure}{.5\textwidth}
\centering
\includegraphics[width=.95\linewidth]{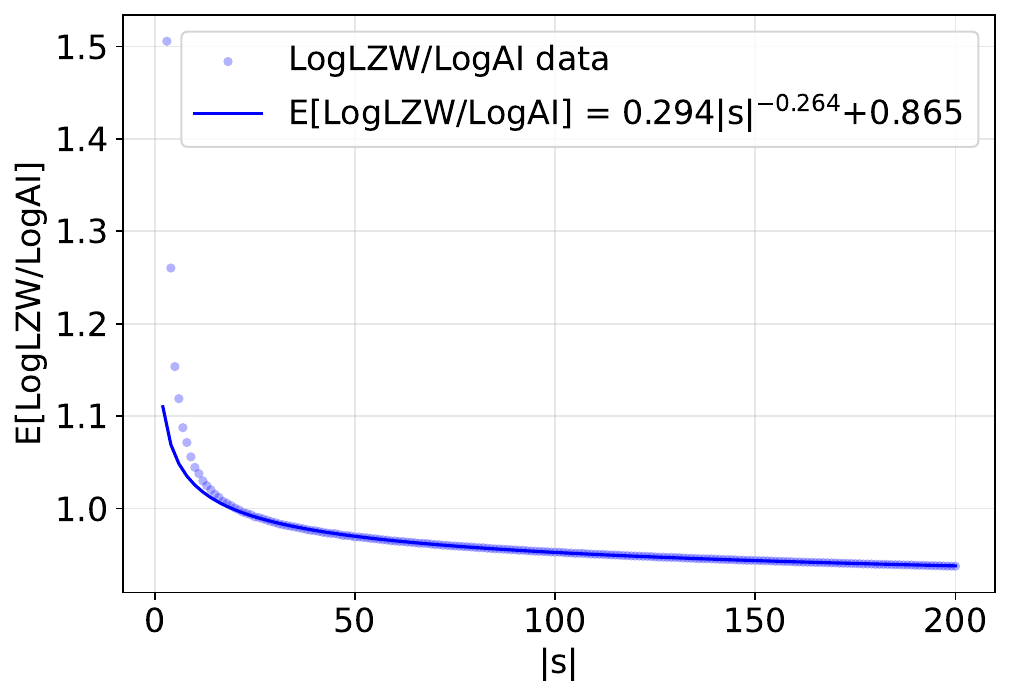}
\caption{}
\label{fig:sub12}
\end{subfigure}%
\begin{subfigure}{.5\textwidth}
\centering
\includegraphics[width=.98\linewidth]{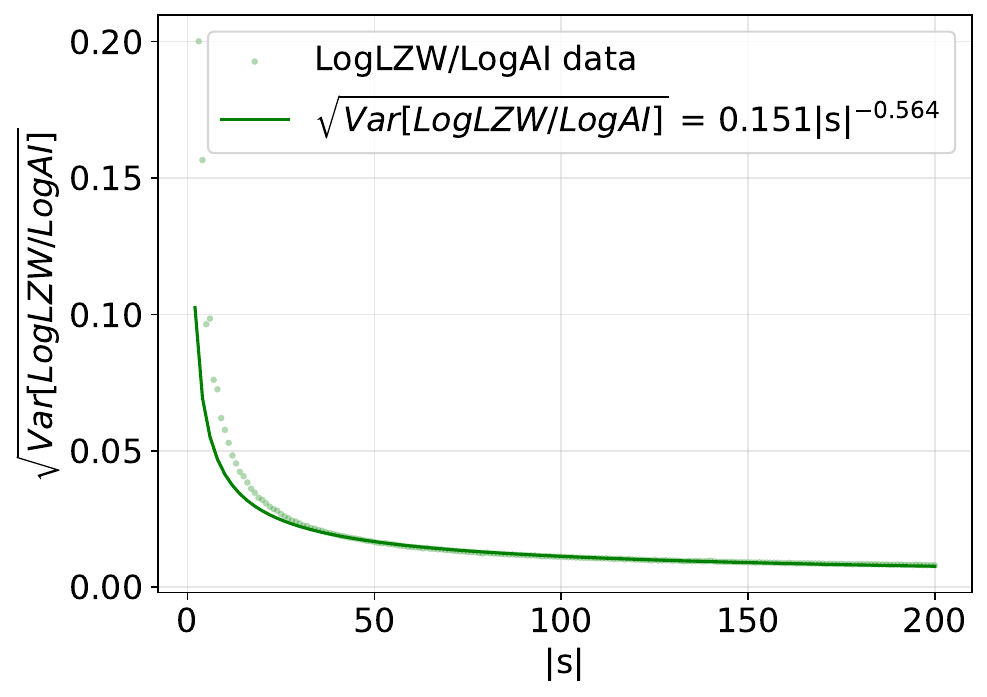}
\caption{}
\label{fig:sub22}
\end{subfigure}
\caption{Asymptotic behaviour of (a) the mean and (b) the standard deviation of the ratio between log LZW and log Ai values for strings up to size 200 built from randomly picking characters from the ZBC string 10,000 times for each size. These plots quantify the Assembly index (Ai) full convergence to LZW.}
\label{fig:asymptlog}
\end{figure}

The linear asymptotic behaviour of the long string for Ai and LZW is shown in Fig.~\ref{fig:asymptlog} indicating how the mean and standard deviation of the ratio of the logarithms of LZW and Ai behave and correlate. In~\cite{abrahao2024}, however, we had already proven the mathematical equivalence the authors did not address.

Tables~\ref{tabpatterned} and~\ref{tabrandom} present the evolution of Spearman correlations for the growth of patterned and random sequences of strings, respectively.

\begin{table}[htbp]
\centering
\caption{Averaged Spearman correlations of growing patterned sequences (increasing copy number) between assembly index and other complexity metrics that the original authors should have compared with under good scientific practice but have always avoided to. Correlation is true for short strings, even with under 15 copies of the same element (the threshold for life, according to AT). The Block Decomposition Method (BDM), was introduced almost a decade before AT. Unlike the claim in~\cite{kempes2024}, the assembly index correlation does not weaken with object size because AT with Ai instantiates Shannon entropy through LZ. However, BDM has been proven to go beyond Shannon entropy and therefore beyond Ai. However, it takes as a baseline what Ai and Shannon entropy do~\cite{abrahao2024} which is the `counting repetitions' baseline that is the most trivial and considered by virtually every other complexity measure~\cite{zenilbdm}.}
\begin{tabular}{cccc}
\toprule
\textbf{Length} & \textbf{LZW} & \textbf{Entropy Rate} & \textbf{BDM} \\
\midrule
8 & 1.00 & 0.76 & 0.99 \\
14 & 1.00 & 0.62 & 1.00 \\
20 & 1.00 & 0.71 & 1.00 \\
40 & 1.00 & 0.85 & 1.00 \\
60 & 1.00 & 0.90 & 1.00 \\
80 & 1.00 & 0.93 & 1.00 \\
100 & 1.00 & 0.94 & 1.00 \\
200 & 1.00 & 0.97 & 1.00 \\
500 & 1.00 & 0.99 & 1.00 \\
1000 & 1.00 & 0.99 & 1.00 \\
3000 & 1.00 & 1.00 & 1.00 \\
\bottomrule
\end{tabular}
\label{tabpatterned}
\end{table}

\begin{table}[ht]
\centering
\caption{Convergence of averaged Spearman correlations of growing random sequences between assembly index (Ai) and various other complexity measures. Not surprisingly, because LZW, Entropy rate and BDM count exact copies in data as their most basic test for nonrandomness, the convergence is fast and full. Ai is the baseline of these other algorithms. In addition, BDM also accounts for linear and algorithmic transformations (reversion, complementation, inversion, etc.)~\cite{zenilctm,zenilbdm}.}
\begin{tabular}{cccc}
\toprule
\textbf{Length} & \textbf{LZW} & \textbf{Entropy} & \textbf{BDM} \\
\midrule
20 & 0.88 & 0.67 & 0.91 \\
40 & 0.90 & 0.83 & 0.95 \\
60 & 0.92 & 0.89 & 0.97 \\
80 & 0.95 & 0.92 & 0.98 \\
100 & 0.95 & 0.94 & 0.99 \\
200 & 0.97 & 0.97 & 1.00 \\
500 & 0.98 & 0.99 & 1.00 \\
1000 & 0.99 & 0.99 & 1.00 \\
3000 & 1.00 & 1.00 & 1.00 \\
\bottomrule
\end{tabular}
\label{tabrandom}
\end{table}

\begin{table}[htbp]
\centering
\caption{\label{table3}Best distribution fit for every complexity index. All indexes but BDM with a long (Pareto) tail distribution are distributed the same way in the typical Shannon entropy and LZW Bernoulli distribution, in particular, the assembly index. Here 10,000 random ZBC strings of varying length. In contrast, BDM's best fit was a Pareto distribution with the following values for permutations and growing (repeated sequence) cases: (1.2e+08, -6.9e+10, 6.9e+10), D = 0.57, and (1.4e+08, -8.6e+09, 8.6e+09), p-value = 5.8e-85.}
\begin{tabular}{l|l|c|c}
\small{\textbf{Measure}} & \small{\textbf{Distribution}} & \small{\textbf{Permutations}} & \small{\textbf{Growing}} \\
\hline
LZW & Bernoulli & 5.3e+02 & 2.1e+02 \\ \hline
Shannon entropy & Bernoulli & 1.5 & 1.52 \\ \hline
Assembly Index & Bernoulli & 5.0e+03 & 5.0e+03
\end{tabular}
\end{table}

Our analysis of fitting the BDM values for both the growing ZBC string and its permutations demonstrated that the Pareto distribution provides a significantly better fit compared to the Bernoulli, thanks to its long-tail distribution. For the growing ZBC string, the Mean Absolute Error (MAE) and the Pareto fit were 12.24 and 17.86, respectively, while the Bernoulli fit showed much higher errors of 332.49 (MAE) and 332.68 (RMSE). Similarly, for the permutations of the growing sequence, the Pareto fit errors were substantially lower (MAE: 61.47, RMSE: 83.32) compared to the Bernoulli fit (MAE: 260.24, RMSE: 262.67). These results indicate that BDM is the only measure that shows a Pareto distribution, with its ability to model long-tail distributions, unlike the traditional Bernoulli distribution that Ai, LZW and Shannon entropy display---given their equivalent nature---measuring the same features.

Contrary to the claims in~\cite{kempes2024}, the distribution of the assembly indices only demonstrates its similarity to Shannon entropy and statistical compression and is a result of their equivalency at the level of one-to-one mapping hence leading to equivalent distributions. Table~\ref{table3} shows that the best distribution fit for the LZW, Shannon entropy, and the Assembly index is a Bernoulli distribution for all different permutations. In contrast, our alternative metric BDM follows a Pareto distribution~\cite{zenilbdm}, illustrating AT's convergent behaviour to other statistical complexity measures, including Ai, LZ and Shannon entropy cognates.

\subsubsection{Analysis on Molecular Data}

\begin{figure}[htbp]
\centering
\includegraphics[width=0.9\textwidth]{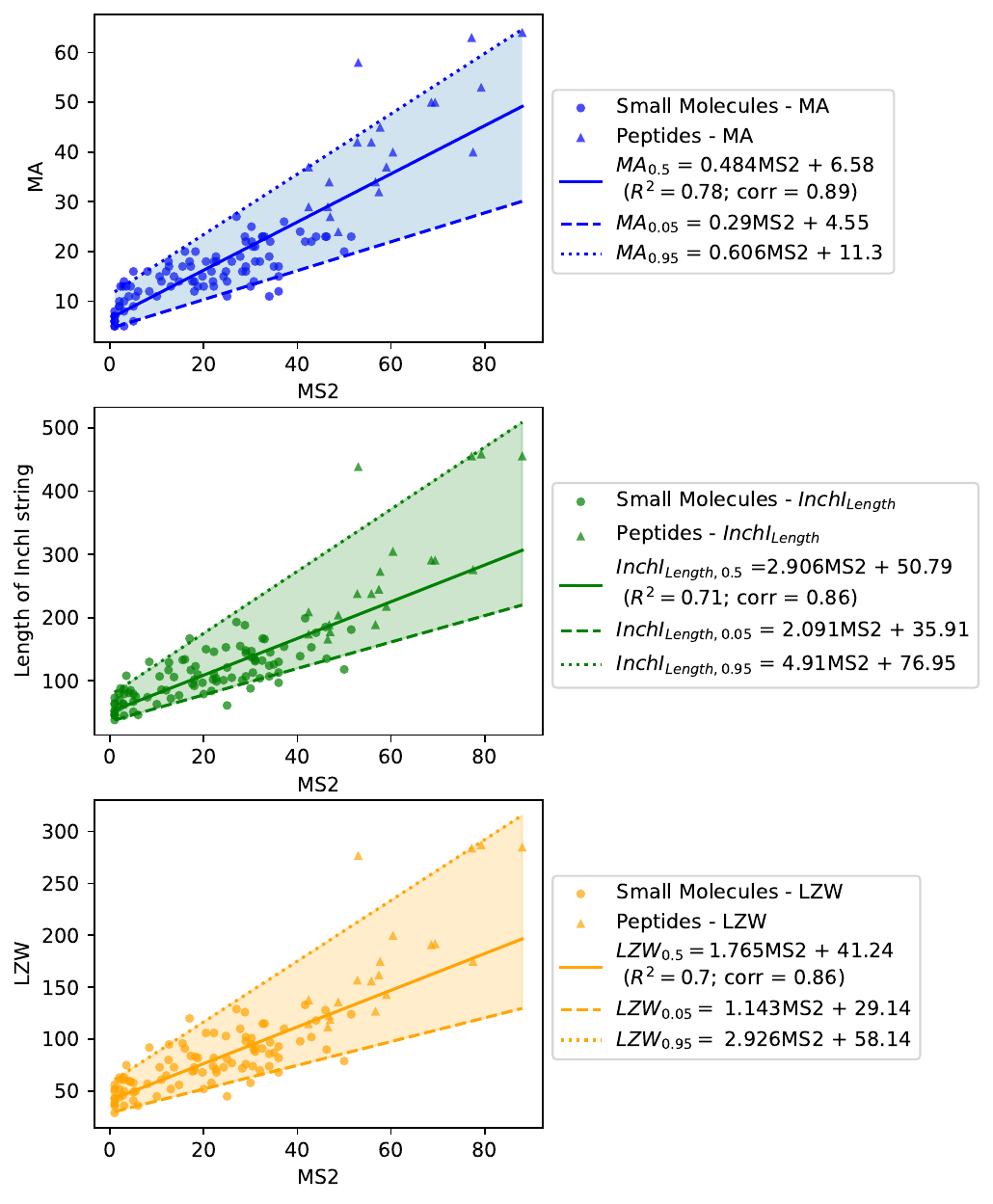}
\caption{Quantile regressions considering 5\%, 50 \% and 95\% quantiles and Pearson correlation (corr) between MA (top), Length of InChI string (middle) and LZW of InChI string (bottom) and MS2 using the data from~\cite{Marshall2021}. The shaded area represents a 90 \% quantile interval. For any of the metrics, linear relationships with similar prediction capabilities are observed. This directly undercuts the AT authors' claims that one of the main advantages of MA (or, equivalently, Ai) is that it could be experimentally measured, since all the other metrics (LZW and the Length of the InChI codes) can also be estimated from experimental MS2 data directly, as we have done here. In contradiction to what the authors claim in~\cite{kempes2024}, the authors of AT have failed at controlling for molecular length, which drives their index. The authors of AT have thus far failed to show a single example in which LZ or Shannon entropy cannot reproduce any results produced by Ai or AT, meaning that Ai and AT are simply equivalent to applying LZ or Shannon entropy with many extra steps.}
\label{fig:AllQuantiles}
\end{figure}

To explain how almost any naive algorithm can reproduce the results reported by the authors of Assembly Theory in~\cite{Marshall2021,Sharma2023}, we took the data from~\cite{Marshall2021} and reproduced their analysis in Fig.~\ref{fig:AllQuantiles} by exploiting the linear relationships found between LZW, the length of the InChI strings for molecules, and MS2 data, respectively.

We say that the length of the InChI string is the number of characters existing after the initial prefix ``InChI=1S/''. We also plot in Fig.~\ref{fig:AllQuantiles} the relation to MA (this is, Ai), as presented by Marshall et al.~\cite{Marshall2021} in their Fig. 3. In such a paper, they call their Ai, `MA', standing for Molecular Assembly; they are for all purposes the same metric based on counting molecular copies. Ai is a generalisation of MA that the authors have applied to objects other than molecules while MA was applied exclusively to molecules. From now on, we will therefore use MA and Ai interchangeably, but will use MA in the context of their first paper applied to molecules from which they claim to be not only connected but able to fully explain selection and evolution.

\begin{figure}[htbp]
\centering
\includegraphics[width=0.7\textwidth]{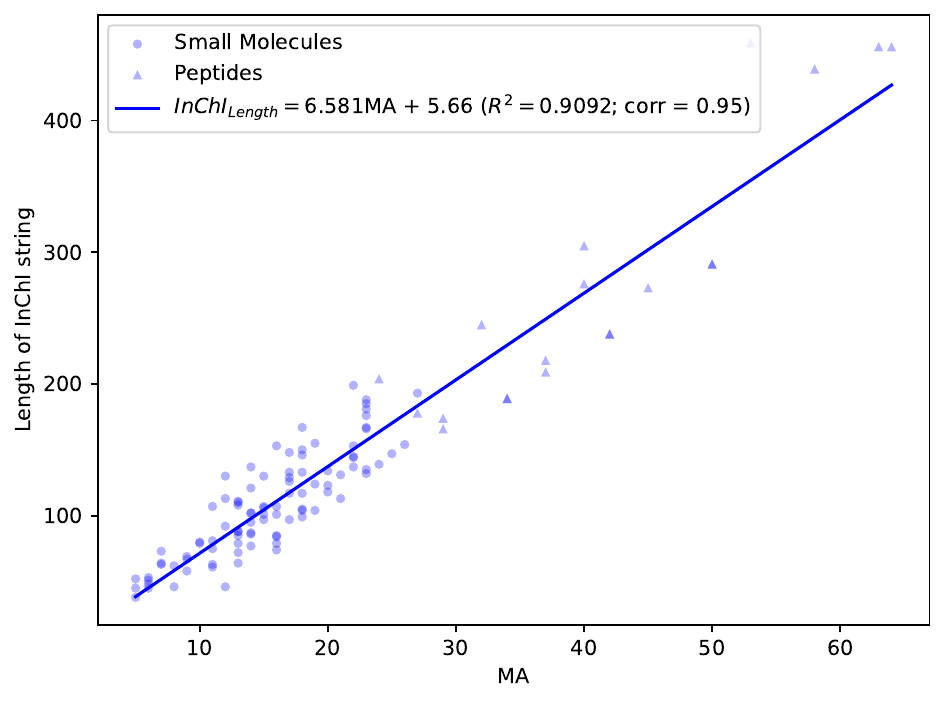}
\caption{Linear relation between MA and the Length of the InChI strings of molecules using the data from the original AT paper~\cite{Marshall2021}. This shows that just by taking the length of SMILES notation, one can reproduce the results of AT and their assembly index. If AT has any advantage over other representations and other compression algorithms, it should produce different if not better results, but it does not. This was conducted on the same set of molecular compounds used by the authors in their paper~\cite{Marshall2021} as the basic control experiment they did not perform, and was reported before to be able to separate organic from non-organic molecules~\cite{zenil2018}.}
\label{fig:MAInChI}
\end{figure}

One might ask why plotting the relation between MA and the length of InchI strings is of interest. This is a consequence of the asymptotic behaviour observed for compression rates in LZ schemes, which tend to a constant for large strings. Since Ai belongs to the LZ family~\cite{abrahao2024}, Ai should be linearly related to the length of the base string for which it is calculated. Thus, it would be natural to observe some type of linear relation, which was verified in Fig.~\ref{fig:MAInChI}.

As shown in~\ref{fig:MAInChI}, the authors of AT did not control for molecular length, which would have deleted any effect of Ai as it does not pick up any other signal.

\begin{figure}[htb]
\centering
\includegraphics[width=0.75\textwidth]{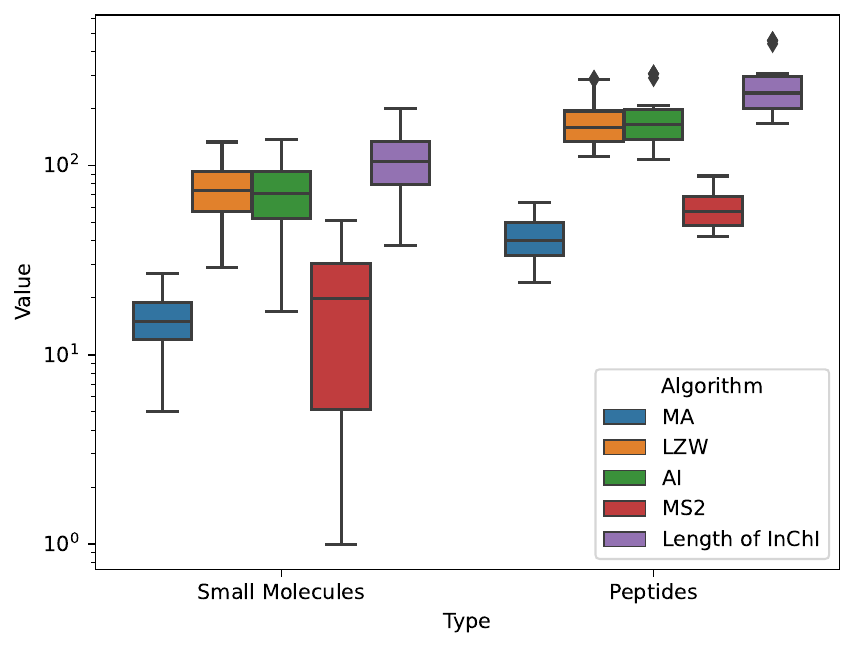}
\caption{Values of several complexity measures for Small Molecules and Peptides, using the data from~\cite{Marshall2021}. This is the calibration experiment performed in~\cite{Marshall2021} before moving to classify organic vs nonorganic compounds. While in~\cite{zenil2018} we normalised by molecular length and compared the efficiency of the organic vs inorganic separation on over 15,000 compounds reported (the full ChemicalData[] database from the Wolfram language), not a small selection~\cite{Marshall2021} of only about 130 compounds, with no comparisons to any other complexity measure that would have disclosed similarity or superiority at separating the classes. MA is one of the many versions of Ai across the authors' papers for application to molecules, implementing their `copy-number' hypothesis.}
\label{fig:groupsMA}
\end{figure}

It is also interesting to see if complexity measures of Small Molecules and Peptides are clustered. This is shown in Fig.~\ref{fig:groupsMA}.

From our numerical experiments it is clear that LZW and Ai produce results that cannot be told apart. In Fig.~\ref{main}, a clear linear asymptotic trend is observed between the LZW and Ai values. This suggests that one can easily derive one from the other except for small local variations that vanish rapidly as a function of length.

Fig.~\ref{fig:asymptlog} shows that the mean value of (log LZW/log Ai) tends to a constant and that the standard deviation of (log LZW/log Ai) tends to 0 for large strings. This indicates that the random variable (log LZW/log Ai) tends to a single-bar histogram, indicating that this ratio tends to a constant without variability. This is an empirical demonstration that the two values converge in alignment with the mathematical proof of equivalence provided in~\cite{abrahao2024}.

Fig.~\ref{fig:AllQuantiles} shows that the claimed link that Ai has with the experimental results is also observed for LZW and even for the length of the InChI strings of the molecules analysed. Thus, if we follow the same rationale as Marshall et al.~\cite{Marshall2021}, it could be said that LZW of the InChI strings, and even on their lengths only, can be experimentally measured by molecules' mass spectra. This lacks, of course, a physical/chemical explanation, which is never provided by AT, beyond circular arguments~\cite{abrahao2024}.

Also, taking into account the results of the experiments in Fig.~\ref{main}, one could investigate how the Ai values are related to the length of the InChI strings. Fig.~\ref{fig:MAInChI} shows a remarkably strong linear relationship, with an $R^2$ of more than 0.9 and a Pearson correlation of 0.95. This indicates that Ai is a proxy for the length of the InChI strings, revealing that these strings already present a great share of the ``concatenation'' information for the molecules.

It is also clear from Fig.~\ref{fig:groupsMA} that not only Ai but also LZW (both applied on the InChI strings), and the length of the InChI strings are reasonably clustered for Small Molecules and Peptides. This further reinforces the deep empirical connection between assembly indices and LZW algorithms.

Regarding the threshold suggested by~\cite{Marshall2021}, that according to the authors separates molecules produced by nonliving systems from molecules produced by living ones, it is claimed that nonliving systems do not produce molecules with MAs (Ai) higher than 15. This has already been found to be erroneous by two other independent groups ~\cite{Hazen2024,raubitzek2024autocatalytic}. However, in their defence~\cite{kempes2024} in this regard, the authors suggest that one has to filter out those examples that prove the Ai hypothesis wrong because they are too artificial, while at the same time, the same authors claim that Ai is completely agnostic and capable of detecting any form of life in the universe~\cite{Marshall2021,Sharma2023}.

\begin{figure}[htb]
\centering
\includegraphics[width=\textwidth]{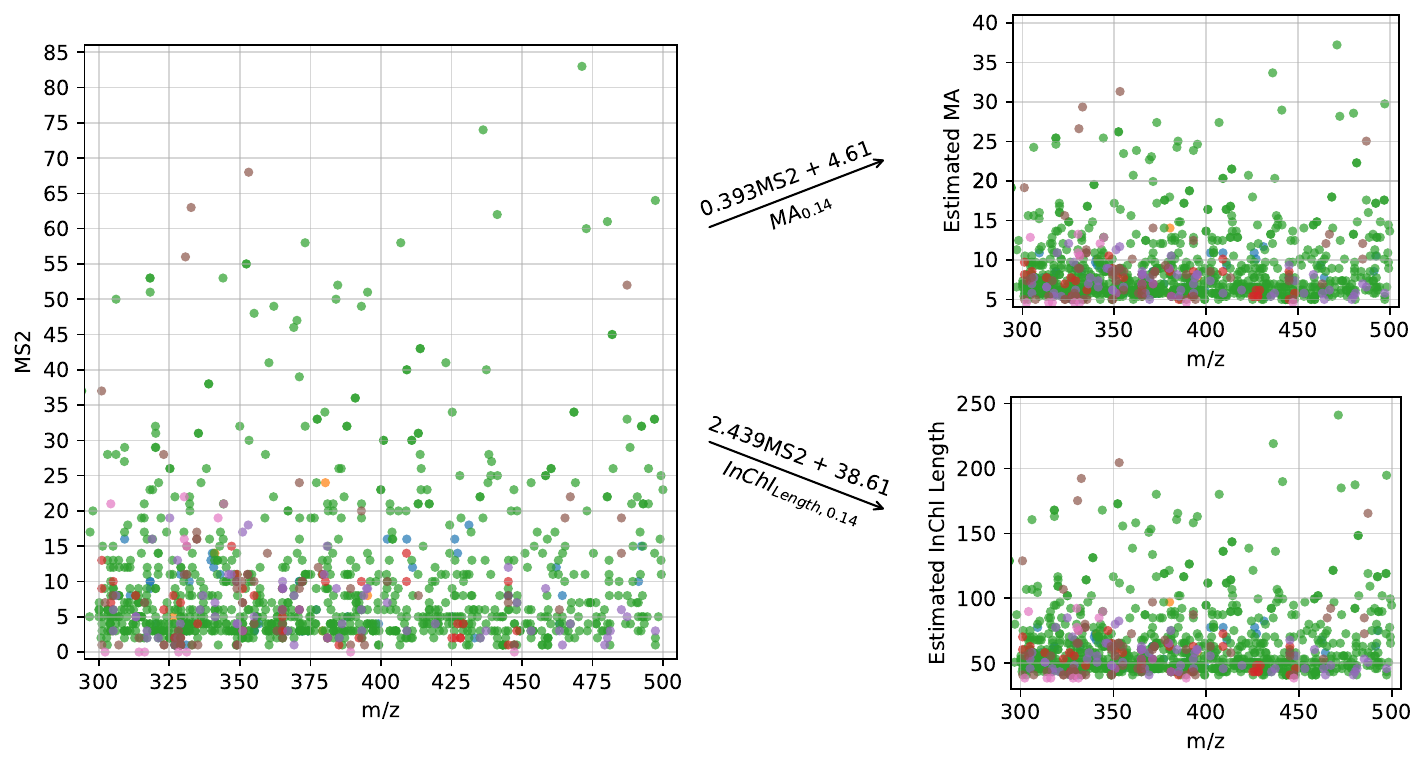}
\caption{Distribution of Measured MS2 peak values, estimated MA (or Ai) values and estimated length of InChI strings versus m/z plotted using the data from~\cite{Marshall2021}. Results are very noisy, with no clear thresholds on a cherry picked set of compounds by the authors of AT in~\cite{Marshall2021}, yet reproducible by molecular length only, not any other features of `life'.}
\label{fig:fig4marshall}
\end{figure}

This claim follows from \cite[Fig.~4]{Marshall2021}, where it is asserted that only molecules produced by living organisms (and some ``Blinded'' samples from NASA/other sources that were not made fully available in their paper's dataset, bypassing Nature's rules and review procedures) have MAs (Ai) higher than 15. By using their own data, we reproduced their \cite[Fig.~4]{Marshall2021} in our Fig.~\ref{fig:fig4marshall}, taking the results for Ai at face value as they were not reproducible, given that despite their claims to the contrary, they did not include all the data. The threshold has been already challenged and found to be in error twice by other independent groups~\cite{Hazen2024,raubitzek2024autocatalytic}, but here we show clearly that their assembly index (Ai) is not picking up any special signal from life.

The first fact worth noting is that Marshall et al.~\cite{Marshall2021} did not use any of the quantile regressions they indicated in their~\cite[Fig.~3]{Marshall2021} to estimate MA (Ai) values from MS2 peaks (i.e., they did not do any of the 5\%, 50 \% and 95\% quantiles, which are also reproduced in Fig.~\ref{fig:AllQuantiles}). By experimenting with the quantile value, it was possible to determine that they used a 14\% quantile regression on the data to build their linear relationship. Then, using this linear relationship and the MS2 peak measurement data, Fig.~\ref{fig:fig4marshall} was obtained. This choice was not explored in depth in their paper, indicating a serious methodological gap.

If we follow the same procedure (build a linear model equivalent to a 14\% quantile regression over the data from~\cite{Marshall2021}), it is also possible to plot the correspondent distribution of estimated InChI string lengths based on MS2 data. This is also presented in Fig.~\ref{fig:fig4marshall}.

If we follow the same rationale as in~\cite{Marshall2021}, we could say that molecules produced by nonliving systems could not have estimated InChI string lengths greater than 100 characters. This illustrates why their claim is neither methodologically robust nor correct. Thus, the analysis of Figs.~\ref{fig:AllQuantiles} and~\ref{fig:fig4marshall} shows that the organic vs inorganic separation claimed in~\cite{Marshall2021} is driven by molecular length and not any other feature captured by their (molecular) assembly index or their copy number, beyond triviality.

We reported all this years before AT in~\cite{zenil2018}, using a more comprehensive experiment involving over 15\,000 compounds, and with proper control experiments by comparing to several measures and several data representations, and without offering the results as proof of unification of physics and biology~\cite{Sharma2023}. The rationale behind Assembly Theory is equivalent to the most common practice of finding any variable linearly correlated to MS2 and then using an arbitrary threshold to separate the molecules, which is without any physical/chemical foundation and does not withstand basic scrutiny.

The authors of AT have unwittingly fallen into the classic correlation as a causation trap, where one finds a correlated variable and claims a causal effect.

Despite the decades of research showing the application of information theory and compression methods to applied sciences, Assembly Theory's proponents keep offering up the fallacy---as in the abstract of~\cite{Jirasek2024Assemblytheory}---that compression methods are not efficiently applicable, or in general that algorithmic complexity-based measures cannot be applied to empirical data or in experimental applications.
In~\cite{Jirasek2024Assemblytheory}, the Molecular Assembly index (based on the assembly index) was shown to correlate (with a Pearson's of 0.88 for their best result) to an ensemble of experimentally measured data.
If this correlation ``demonstrates'' that Assembly Theory is an experimental measure, then it follows that the strong correlation using other metrics we have shown would also ``demonstrate'' that one can measure `assembly' by using LZ or Shannon entropy.

Since the relation between MS2 peaks and MA (Ai) is linear (the same for InChI lengths), what was actually proposed in one of the papers in question~\cite{Marshall2021} is that molecules produced by living and nonliving systems can simply be distinguished by the number of their MS2 peaks (which were measured and not estimated, the authors contend).

\subsection{AT Unsuitability to Quantify Selection or Evolution}\label{sectionevo}

One of the recurring difficulties in engaging with Assembly Theory (AT) is the way its proponents shift domains when their claims are challenged. At the molecular level, the authors of AT insist that the assembly index (Ai) captures the causal essence of biological selection. Yet when it is shown that Ai fails---where Shannon entropy and naive compression already replicate or outperform its results---they retreat to the claim that Ai is broader, extending beyond molecules and even beyond biological evolution. This rhetorical oscillation highlights the absence of a consistent explanatory framework. A measure that cannot withstand scrutiny in its supposed ``home domain'' of molecular complexity cannot be rehabilitated by vague appeals to universality.

Here we offer some arguments showing how AT with its Ai would fail in characterising simple processes of biological evolution and selection. Let $o$ be any object (molecular or not) and $x=\mathrm{enc}(o)\in\Sigma^{*}$ its symbolic encoding.
Let $Ai(x)\in\mathbb{N}$ denote the assembly index of $x$, defined as the minimal number of exact-copy concatenations required to generate it.
Let $g$ be a generative description of $o$, and let the realised phenotype be $\phi(g,o)$.
In an environment $e$, fitness is given by $w(\phi(g,o),e)$.
For a trait $\tau$ with indicator $\mathbf{1}_\tau\in\{0,1\}$, the standard selection coefficient is

\begin{equation}
s_\tau(e) = \mathbb{E}[w(\phi(g,o),e)\mid \mathbf{1}_\tau=1] - \mathbb{E}[w(\phi(g,o),e)\mid \mathbf{1}_\tau=0].
\end{equation}

Trait loss (regressive evolution) occurs when the net benefit $b_\tau(e)$ of $\tau$ falls below its maintenance cost $c_\tau$, so that $b_\tau(e)-c_\tau<0$.

\begin{enumerate}
\item \textbf{Environment dependence.} For fixed $x$ (hereinafter fixed $Ai(x)$), choose environments $e_1, e_2$ such that $s_\tau(e_1)>0$ while $s_\tau(e_2)<0$. Since $Ai(x)$ does not vary with the environment, no function of $Ai(x)$ can uniformly recover the sign of $s_\tau(e)$.

\item \textbf{Trait-loss edits invisible to Ai.} Let $T$ be a causal transformation (deletion, reversion, inversion, simplification) yielding $o' = T(o)$ with encoding $x'=\mathrm{enc}(o')$. Such edits can eliminate costs ($c_\tau \to 0$) without introducing exact-copy patterns. Hence $Ai(x')$ may change arbitrarily, while the true selective dynamics is driven by causal edits invisible to Ai.

\item \textbf{Impossibility.} If there existed $F$ with $s_\tau(e)=F(Ai(x))$ for all $x,e$, then by (1) the same $Ai(x)$ would imply contradictory signs of $s_\tau(e)$. Therefore, no such $F$ exists. Selection coefficients cannot be functions of Ai (or any statistic of isolated encodings) alone.
\end{enumerate}

Whether $o$ is molecular or not, selection---and de-evolution in particular---depends on $(o,e)$, not on $o$ in isolation. Ai, as a function of isolated encodings, cannot capture the environment-dependent and causally structured nature of evolutionary dynamics. Thus, Ai cannot be a good metric for selection or evolution if it cannot explain even its most basic operations.

\subsubsection*{The Cavefish and the Limits of Assembly Theory}

The Mexican blind cavefish (Astyanax mexicanus) provides a concrete biological case that exposes the conceptual and empirical limits of Assembly Theory (AT) and its Assembly Index (Ai). This species has evolved multiple independent cave-dwelling populations derived from surface ancestors, exhibiting striking regressive and adaptive traits such as loss of pigmentation, eye degeneration, expanded taste buds, and rewired sensory systems. These transformations are driven by well-understood evolutionary mechanisms including natural selection, relaxed selection, and environmental constraint—all processes that are inherently causal, reversible, and context-dependent.

Such systems are directly relevant to AT because their authors explicitly claim that their framework `reconstructs the evolutionary tree of life', `captures causal molecular relationships', and `maps evolutionary trajectories' from only molecular data. If AT genuinely encodes causal history and evolutionary change, it should at least accommodate these canonical cases where adaptation and trait loss are understood in molecular and developmental terms. The cavefish therefore serves as a clear empirical test for AT's central claim.

\begin{table}[ht] \centering \scriptsize \caption{Evolutionary mechanisms in cavefish and the limits of Ai/AT, contrasted with the broader capabilities of BDM/CTM.} \begin{tabular}{p{2.6cm}p{3cm}p{3.4cm}p{3.4cm}} \toprule \textbf{Evolutionary mechanism} & \textbf{Biological example} & \textbf{Why Ai/AT fails} & \textbf{How BDM/CTM may capture the traits} \\ \midrule Environment-dependent selection & Eyes are adaptive in light-rich rivers, but maladaptive in dark caves & Ai assigns a fixed value to molecular encodings, blind to environmental context & BDM/CTM incorporate environment via algorithmic probability and reweighting of generative models \\ \midrule Regressive evolution (trait loss) & Eye reduction and disappearance saves energy in darkness & Trait loss involves deletions and simplifications, not repetitive motifs (even at the molecular level) & BDM accounts for deletions, reversions, and inversions; CTM explores alternative generative programs reflecting simplification \\ \midrule Selection not reflected at molecular level & Enhanced taste buds, mechanosensory cells, and altered foraging behaviours & Adaptive traits arise through regulatory/neural changes, invisible to Ai's copy-counting & BDM captures causal, non-statistical regulatory patterns; CTM identifies minimal generative rules beyond copy-number \\ \bottomrule \end{tabular} \end{table}

The cavefish example demonstrates that evolution is not a monotonic accumulation of structural novelty or statistical regularity. It involves gains, losses, and shifts in function that depend on the ecological context and developmental regulation. Ai reduces all these processes to an invariant count of combinatorial steps, blind to the mechanisms by which traits appear or disappear. Even at the molecular level, most adaptive changes in Astyanax occur through altered gene expression and sensory integration rather than through the creation of new metabolites. Consequently, any approach restricted to molecular assembly complexity would miss precisely the causal signals it claims to capture.

Moreover, when AT is applied to actual molecular datasets, its results are indistinguishable from those produced by simple surrogates such as Shannon entropy, GC content, or compression-based similarity. These established information-theoretic measures already reconstruct molecular evolutionary trees with high accuracy and known statistical properties. If AT cannot outperform these benchmarks, its claim to causal or evolutionary insight collapses into rebranded statistics. 

The cavefish illustrates the broader conceptual failure of AT's central premise. The theory conflates structural description with evolutionary causation, ignoring the roles of environment, regulation, and developmental plasticity. If Ai cannot account for the causal mechanisms that drive adaptation in such a textbook case of evolution, then its promise to explain selection and evolution from molecular data alone cannot be sustained. The example therefore stands not as an isolated counterpoint but as representative of the many causal, non-trivial features of evolution that AT is structurally incapable of detecting.

Ai reduces selection and evolution to trivial statistics, missing the causal mechanisms at play that do not lead to statistical clues not reflected in the static outcome of the object at a given time, despite the many suggestions that AT and Ai capture the dynamic causal history of the object.

This example undermines AT's pretension of generality or relevance in the very core area in which they have introduced it. Even at the molecular level, Ai performs no better than simple surrogates, as we have shown, the raw length of molecular encodings such as InChI strings and the application of Shannon entropy to their own data reproduces their results. If Ai fails in its core claim of detecting selection from molecular data, extending it before and beyond molecules, but it cannot reproduce any new result, it cannot be presented as a new theory but rather as a pedagogical exercise.

In contrast, measures grounded in algorithmic information theory, such as the Block Decomposition Method (BDM), introduced before AT, subsume trivial copycounting while are also capable of capturing other causal operations ubiquitous in biology such as deletions, reversions, inversions, and other structured mechanistic transformations.

\subsection{Block-decomposing the Assembly Index}\label{sectionBDM2}

As detailed in Section~\ref{sectionMethods}, the Block Decomposition Method (BDM) offers a more robust framework for complexity estimation than the assembly index (Ai). By its construction, Ai can never perform better than BDM because Ai's core mechanism---counting exact copies---is merely the baseline test for nonrandomness that BDM incorporates.

As discussed in Section~\ref{sectionMethods}, we demonstrated in~\cite{abrahao2024} that the LZ parsing of any object according to the assembly index (Ai) calculation method is given by the longest (rooted) assembly paths that belong to the minimum rooted assembly subspace from which the assembly index of the object is calculated.

Now, suppose a sufficiently long string $x$ is generated according to a \emph{stochastic source} for which the noiseless source coding applies~\cite{Cover2005}, e.g. one following an i.i.d. probability distribution.

Then, it is known that the best \emph{statistical} compression rate by any method is lower bounded by entropy-based compression. Statistical compression methods such as Huffman, LZW or LZ78 are universal in that can they asymptotically all converge to Shannon entropy ($\mathbf{H}$) rate limit~\cite{Cover2005}.

Therefore, from Equation~\eqref{eqBDMbounds}, there is a computable---in the case of LZW, also computationally efficient---method $m$ and a partition $i$ of sufficiently small blocks such that
\begin{equation}\label{eqBDMandAi}
BDM( x , i , m' ) \leq \mathbf{O}( 1 ) + N_i \mathbf{H}_i\left( X^i \right) \leq \mathbf{O}( 1 ) + K_{Ai}\left( x \right) \text{ ,}
\end{equation}
where $K_{Ai}(x)$ is the size in bits of the compressed form according to the assembly index calculation method.

In the \emph{general case}, one can also directly demonstrate that the assembly index calculation method can never outperform BDM for any generative process (as opposed to simple stochastic sources), whether one has a fully deterministic source, stochastic source, or a mixed process. We have demonstrated in~\cite{abrahao2024} that any a priori chosen encoding scheme $\left< V\left( {\Gamma^*}_y \right) \right>$ of the set $V\left( {\Gamma^*}_y \right)$---an encoding which is agnostic not only vis-a-vis the choice of the final object, but also the (sub)space structural representation---of vertices of the minimum rooted assembly subspace ${\Gamma^*}_y$ of the (final) object $y$ losslessly compresses/encodes the object $y$. Thus, there is a partition $i$ sufficiently close to $i_0$, and there is an encoding-decoding scheme $m$---under computational resources of the same order as those expended to calculate the assembly index---such that
\begin{equation}\label{eqBDMandAicomputablesource}
BDM( x , i , m ) \leq \left| \left< V\left( {\Gamma^*}_y \right) \right> \right| + \mathbf{O}(1) = K_{Ai}\left( x \right) \text{ ,}
\end{equation}
where $|\left< V\left( {\Gamma^*}_y \right) \right>|$ is the size in bits of the arbitrarily chosen encoded form of the set $V\left( {\Gamma^*}_y \right)$. Thus, Ai is a compression method subsumed within BDM (see also Equation~\eqref{eqMain}).

From Equation~\eqref{eqBDMandAicomputablesource}, and fundamental properties in AIT such as the algorithmic coding theorem and those discussed in Section~\ref{sectionBDM}, we also have it that
\begin{equation}\label{eqKandAicomputablesource}
\begin{aligned}
- \log\left( \sum\limits_{\mathbf{U}(p) = x} \frac{1}{2^{|p|}} \right) &= \\
\mathbf{K}(x) \pm \mathbf{O}(1) &\leq \\
BDM(x, i_0, m) \pm \mathbf{O}(1) &= \\
CTM(x, Q) \pm \mathbf{O}(1) &\leq \\
K_{Ai}(x) \pm \mathbf{O}(1) & \text{ ,}
\end{aligned}
\end{equation}
where $\sum_{\mathbf{U}(p)=x} 2^{-|p|}$ is the \emph{universal a priori probability}---or, equivalently, the \emph{algorithmic probability} (AP)---of the event $x$.

Contrary to multiple claims by the authors of Assembly Theory and their examples~\cite{kempes2024}, we have empirically shown the asymptotic behaviour of Ai to other metrics, including LZW.

Ai offers no advantage over BDM, Shannon entropy ($\mathbf{H}$) or LZW, but also no time complexity advantage, as its calculation, according to the authors themselves, is intractable (a fact they present as an advantage)~\cite{kempes2024}. For this reason, they have also had to come up with different heuristics in different papers~\cite{Marshall2021,Marshall2022TheoreticalAT,Sharma2023}.

The following equation provides a hierarchical relationship between complexity methods to locate the place of each metric. Let $A \preceq_{cp} B$ denote the (partial) order of a method $B$ subsuming another complexity method $A$ (and $\equiv_{cp}$ when both $A \preceq_{cp} B$ and $B \preceq_{cp} A$ hold). Thus, from~\cite{abrahao2024} and Equations~\eqref{eqBDMbounds},~\eqref{eqBDMandAi},~\eqref{eqBDMandAicomputablesource}, and~\eqref{eqKandAicomputablesource}, one has it that the following relationship
\begin{equation}
\textnormal{Ai} \equiv_{cp} \textnormal{LZ} \preceq_{cp} \mathbf{H} \prec_{cp} \textnormal{BDM} \preceq_{cp} \textnormal{CTM} \prec_{cp} \textnormal{AP} \equiv_{cp} \mathbf{K}
\label{eqMain}
\end{equation}
holds. Where Ai and $\mathbf{K}$ are at the two extremes of the complexity spectrum~\cite{Chaitin2004} and Ai on the trivial side, weaker or equal to popular statistical compression and Shannon entropy as only counting repetitions, while computable approximations related to stronger measures have been proven to better capture causality~\cite{iscience} related to selection and evolution~\cite{hernandez2018}.

In addition, BDM is computable given the relaxation of the condition of not having to find or prove program length minimality (or, equivalently, all the generating programs). Ai is the weakest metric, unable to capture any nontrivial pattern, let alone nontrivial causal content; therefore, it cannot explain selection or evolution despite its circular arguments~\cite{abrahao2024}, especially not any better than LZ and Shannon entropy can~\cite{Uthamacumaran2024}.

Equation~\ref{eqMain} summarises some of the most important historical and methodological concepts in the history of complexity theory, creating a hierarchical structure of complexity measures into which Ai was never properly inserted, the senior authors even suggesting that Ai figured at the rightmost end as a generalisation of all other complexity measures~\cite{fridman2024}. Not only is that wrong, with Ai arriving probably 50 years late, but Ai implements the weakest form of test that any other measure of complexity would already take as its baseline.
In fact, the same copy-number argument has been used in the same context for decades, even in the reconstruction of evolutionary trees~\cite{cilibrasi,zenilreview} and in the addressing of some of the greatest challenges in biology, such as nucleosome positioning~\cite{zenilnar} in connection with GC content (counting nucleotides G and C). We made this case in a set diagram in~\cite{abrahao2024}, here formalised in Equation~\ref{eqMain}. Every measure formulated to date can be inserted somewhere between these lower and upper bounds.

\section{Discussion}\label{sectionDiscussion}

Although Assembly Theory has been promoted in some outlets as a scientific breakthrough, driven in part by press releases, it has yet to demonstrate substantive contributions beyond established concepts in complexity and information theory. This paper, in continuing to refute hyperbolic claims made on behalf of AT, serves to correct several misconceptions about computability, algorithmic complexity, and compressibility.

Our results confirm that an object with a low assembly index has high LZ compressibility and therefore low Shannon entropy, and an object with a high assembly index has low LZ compressibility and therefore high Shannon entropy. AT and its assembly index can therefore only reproduce the capabilities already known for entropy-based metrics, for which LZ algorithms are already optimal estimators.

To the extent that AT may have a pedagogical value in its peculiar application to molecular complexity, this has unfortunately remained unclear due to the hyperbole surrounding the introduction of AT by its authors touted as a theory capable of even explaining the expansion of the universe and cosmic inflation~\cite{glasgow,asu,templeton}.

It may appear to be an advantage of AT that it appears as a transparent-box algorithm. This is in fact not the case, since a different heuristic of the same metric has been introduced in every new paper. Most of the time, this forces us to take their empirical results at face value, thus giving them the greatest advantage while hindering reproducibility. Moreover, one can always open the box of a statistical compressor and force the definition of dictionary elements to make the compressor fully transparent.

We have been very critical of the use and abuse of compression as a complexity measure before, and we have been addressing this issue for more than a decade~\cite{zenilreview}. When it comes to Assembly Theory, we think that the limitations of statistical compression cannot be rectified with another compression algorithm of exactly the same limited nature.

In this paper, we are not advocating for statistical compression~\cite{zenilreview}. For more than a decade we have worked to move the field away from statistical compression and simplistic measures of complexity like AT towards recursive ones capable of more than statistics, like CTM and BDM.

We recognise the effort of the authors of AT to offer a more formal framework for discussion of the details relating to their index in~\cite{kempes2024}, unlike previous attempts~\cite{Marshall2022TheoreticalAT} that lacked any attempt to compare and engage with current literature and methods, moving away from other previous mistakes made by the same senior authors, ranging from providing an unnecessary proof of computability~\cite{Marshall2022TheoreticalAT} to baseless claims that AT and Ai were a generalisation of algorithmic complexity (Solomonoff-Kolmogorov-Chaitin) as voiced in~\cite{fridman2024}. However, the discussion around AT has historically been plagued by hype and obfuscation since its inception.

\subsection{Unfounded new arguments based on stochasticity}

One of the authors of AT's objections is that compression is a modern concept from computer science, hence anachronistic vis-\`a-vis the prebiotic phenomena, computers not having predated humans, they say. This argument makes no sense from any perspective. As scientists we are meant to analyse nature with the best tools humans have at their disposal, whether drawn from mathematics or computer science. One does not need to justify mathematics as human-made to apply it as a universal tool. Moreover, the core concepts of AT such as `copy', `number', or `graph' are also human constructions and do not predate biotic systems on Earth either.
The authors of AT tend to confound the way an algorithm can be described with how nature may instantiate it~\cite{abrahao2024}. Describing a hypothesis is not the same as explaining the phenomenon for which they have proposed a hypothesis, especially given Assembly Theory's poor performance compared to `a-vis Shannon entropy or LZW offering no advantage.

For every assembly (sub)space there is a compressing grammar that not only generates the same objects but also is isomorphic vis-a-vis the particular multigraph put forward by Assembly Theory~\cite{abrahao2024}.

In fact, because all of the methods of Assembly Theory are computable (even if the calculation of the assembly index may be intractable), there is an unrestricted grammar that generates---via a computable process---all the possible assembly spaces.

Arguments around the unsuitability of Turing machines or algorithmic complexity to deal with uncertainty and stochasticity~\cite{Wolpert2024StochasticProcessTuring,kempes2024} overlook the fact that algorithmic probability is the accepted mathematical Bayesian theory of optimal induction/abduction. We elaborate more on this in~\ref{sectionPhysicalGrounding}.

Furthermore, Shannon entropy is the quintessential metric to deal with stochasticity, which is equivalent to AT and its assembly index, therefore, making redundant the use of another metric~\cite{abrahao2024} to deal with the same challenge.

\subsection{Unfounded new arguments based on Time complexity}

Another seminal flaw in the latest arguments from AT's authors~\cite{kempes2024} lies in their reliance on inconsistent heuristics, which change from paper to paper. These heuristic tweaks, designed to maintain marginal differences from Shannon entropy or LZ compression, amount to shifting goalposts. Despite theoretical and empirical proofs demonstrating that the assembly index is fundamentally equivalent to Shannon entropy, the authors persist in claiming novelty by pointing to small scatter deviations in the data. However, these deviations arise not from theoretical innovation but from arbitrary algorithmic inefficiencies.

Although time complexity is an important feature of an algorithm in computer science, in science, and for scientific purposes, the strength of a method lies in its explanatory power and predictive accuracy, not in its inefficiency or efficacy. However, the new argument of the AT group is that they are worse and slower and, therefore, are different. Although this argument borders on sounding absurd as a positive attribute, it is also wrong. There are already slow implementations of LZ algorithms that may converge faster or not to Shannon entropy rate. Even compression tools like gzip have compression levels as input parameter ranging from 1 to 9, with 1 being the fastest compression (but with a lower compression ratio) and 9 being the slowest compression (but with the highest compression ratio) depending on variations such as multiple passes through the object.

The AT authors' new argument based on time complexity leads to a contradiction. They dismiss Kolmogorov complexity for its reliance on the Turing machine formalism which it does not, yet appeal to time complexity---a framework entirely dependent on that same formalism---as a new defending argument in favour of Ai.


Time complexity theory provides a powerful framework for classifying algorithms under the Turing machine model, but its relevance to natural processes is minimal. Complexity classes such as P, NP, and PSPACE are entirely model-dependent and assume the sequential execution of operations under the conventions of the Turing formalism. Even minor changes to the computational substrate can radically alter time-complexity results. For example, adopting a parallel architecture such as a cellular automaton can collapse exponential processes into linear or constant-time ones. Unlike algorithmic (Kolmogorov) complexity, which remains invariant up to an additive constant across all computational models, time complexity does not survive architectural changes.

A simple physical example illustrates this disconnect. Consider the task of sorting straws by length. In standard algorithmic analysis, comparison-based sorting is bounded below by $O(n\log n)$ operations. In the physical world, however, one may grasp a bundle of straws and tap them against a table, causing them to align by length in a single action---effectively an $O(1)$ sorting procedure. This highlights that time complexity classes are artefacts of a particular abstract model, not descriptions of physical resource use. Physical systems exploit parallelism, dynamics, and material constraints that are not captured by asymptotic complexity classes.

The contradiction in the defence of Assembly Theory (AT) and the assembly index (Ai) is therefore stark. AT mistakenly dismisses Kolmogorov complexity on the grounds that it relies on the Turing machine formalism, while simultaneously appealing to time complexity classes to distinguish Ai from compression. Yet, by virtue of requiring Turing universality, unlike time-complexity, Kolmogorov complexity is independent of the formalism or architecture of the universal constructor and its Invariance Theorem, while time complexity is entirely dependent upon it. Invoking time complexity as a defence of Ai while rejecting Kolmogorov complexity is thus not only misplaced but also reveals a repeated conflation and misunderstanding of the foundations of these areas within AT.

\subsection{False arguments from incomplete experiments}

We have demonstrated that to separate molecules produced by living and nonliving systems, Marshall et al.~\cite{Marshall2021} used a linear model to estimate MA (Ai) values from MS2 (mass spectral) peak data. Then, we showed that this linear model (which is the core of the methodology in~\cite{Marshall2021}) is easily replaced by any other model that correlates any metric with the MS2 peak data.

The authors of Assembly Theory are mistaken in claiming to have been the first to report on the separation of organic from inorganic compounds using complexity measures. We reported this separation in 2018~\cite{zenil2018} while they did so in 2021~\cite{Marshall2021} (on a small set of molecules). In 2018, we also reported connections to selection and evolution~\cite{hernandez2018}, which they did in 2023~\cite{Sharma2023}. All this without the hyperbolic claims that have unjustifiably followed Assembly Theory.

We have shown here that it is possible to build a linear model, with similar predictive performance, to estimate the InChI string lengths of the molecules from MS2 peak data. 

Although some complexity measures may be similar or may overlap in scope or method, as stated in~\cite{kempes2024}, one has to demonstrate any added value or justifiable advantage in the predictive or explanatory power of the explanation.
Assembly Theory is neither predictive nor explanatory. The results shown so far are easily reproducible or outperformed, and the alleged explanatory power is based on very empirical results that do not indicate that their peculiar assembly processes are in any way favoured. The connections to selection or evolution are only speculative and are based on circular arguments~\cite{abrahao2024}. We further provide more examples in this work.

It cannot be expected that AT will provide any insight into how nature or the universe works that cannot be achieved using other traditional measures already existing that are based on exactly the same ideas and principles. AT does not offer any new predictive or explanatory power that could justify claims that AT can unify physics and biology, characterise life on earth and the universe, or revolutionise medicine and drug discovery~\cite{glasgow}.

We have not yet found a domain in which Assembly Theory explains or produces results that cannot be replaced by some other naive statistical algorithm replicating or outperforming the results of Assembly Theory~\cite{zenil2018,hernandez2018}, all of which cast doubt on the utility of AT and its assembly index.

\subsection{The Assembly index as a metric}

The authors of Assembly Theory suggest that their measure has not been correctly understood as a metric because their assembly index is a measure able to quantify complexity, selection, and evolution. This seems to imply that other measures of complexity do not, from Shannon entropy to Lempel-Ziv (LZ) complexity widely used as metrics for quantifying complexity for decades ranging from spam detection to phylogenetic reconstruction. These metrics are well-established, rigorously tested, and have contributed to scientific understanding. The claim that AT introduces something novel in this domain that can quantify selection and evolution is not supported by any evidence. This is especially damning given that we have demonstrated both empirically and mathematically that their results can be fully reproduced using naive statistical tools such as Shannon entropy and LZ compression, and that their thresholds for life can be also reproduced by these other traditional tools. However, the authors have consistently refused to perform control experiments that would compare their measure with established alternatives without providing any proof of superiority.

What AT has yet to demonstrate is that the assembly index can achieve results that other measures cannot, particularly in relation to selection and evolution, but the authors have never experimentally compared their measure to any other index, and their strategy has been to dismiss these other measures from the onset. If it is true that AT offers a better, or even a different, explanation of evolution, selection, or life than Shannon entropy, the authors must prove it. So far, AT has offered neither a different explanation nor better predictions than those offered by Shannon entropy metrics and LZ compression, as we have shown~\cite{abrahao2024}.

The authors of AT have suggested that the statistical algorithms we used in previous articles~\cite{Uthamacumaran2024,abrahao2024} that successfully reproduced all of their results did so because we did not control for molecular length. However, this was precisely our original criticism of their methods. Our findings show that their assembly index is entirely dominated by molecular length, as we have demonstrated in this contribution, proving that their separation between organic and nonorganic compounds was determined exclusively by length.

Our argument has always been that the authors of AT did not control for the most basic features to compare their index with other measures of complexity. Had they done so, they would have immediately found that their whole groundbreaking theory was unable to produce any new results, that the results were entirely driven by Shannon entropy and compression techniques.

The authors of AT also used the argument that we are suggesting that statistical data compression can characterise life. Our argument is that none of these algorithms can, including Shannon entropy and statistical compression of which Ai is a derived index and another statistical compression algorithm.

The authors suggest that Shannon entropy and compressibility are not relevant ``metrics'' in their domain. This is incorrect and reveals a lack of engagement with decades of empirical work using entropy and LZW in fields ranging from spam detection to genomics and evolutionary analysis.

\subsection{The illusion of physical grounding}\label{sectionPhysicalGrounding}

Unlike the claims made by the authors of AT in~\cite{kempes2024}, Kolmogorov complexity does not require the construction of a universal Turing machine (UTM). The machine only contributes a fixed additive constant that is discounted in all cases. By contrast, the assembly index is also computed symbolically: it is implemented as code running on a digital computer, itself a universal machine, without anyone needing to build the computer from scratch. If their argument were applied consistently, they would face the same question they pose to Kolmogorov complexity: Who or what is assembling the blocks? The answer, whether ``nature'', the ``laws of physics or biology'', or ``the computer'', is the same in all cases.

The attempt to dismiss Kolmogorov complexity on the grounds that it presupposes a universal Turing machine rests on a basic misunderstanding of the measure itself. In algorithmic information theory, the choice of universal Turing machine is irrelevant beyond a fixed additive constant. The UTM was introduced precisely to demonstrate that one need not account for the particular machine at all---the concept of universality guarantees that the substrate does not matter, so long as it is computationally complete. This is why it has long been possible to speak meaningfully about the Kolmogorov complexity of DNA sequences, molecular graphs, or any other object: the definition applies to the description, not to the physical act of building either the object or the machine.

The historical trajectory further undermines their claim. Turing's universal machine was not conceived as an abstraction detached from reality but as a formalisation of the mundane practice of human ``computers'' calculating step by step with paper and pencil. The abstraction was then added to show that the details of the substrate were immaterial. To invert this order, demanding physical instantiation of the abstraction, is to miss both the logic and the foundations of computability theory.

Invoking von Neumann to demand prior physical construction also misconstrues von Neumann's project. Von Neumann's ``universal constructor'' was an existence proof carried out entirely in a formal medium. The aim was to establish the logical sufficiency of a description-directed constructor for self-reproduction and open-ended construction within an abstract substrate---not to require that such a device be physically built before one can reason about construction or complexity.

\subsection{Physicality v Digital Representations}

One of the most persistent rhetorical devices in this defence is the claim that AT operates on ``physical data'', whereas compression algorithms are said to deal only with ``abstractions''. This narrative has been used to elevate AT as fundamentally different from, and supposedly superior to, established information-theoretic methods. However, this distinction is both intellectually confused and technically incorrect.

The rhetoric of ``physicality'' obscures the fact that both their assembly index and Kolmogorov complexity operate in the same way: as symbolic measures, implemented in computer programs running on universal machines, applied to encoded descriptions of objects rather than to the physical objects themselves. Ultimately, both Kolmogorov complexity and the assembly index are operationalised as algorithms executed on universal machines acting upon symbolic encodings, and Assembly Theory may be understood as a restricted instance of Kolmogorov complexity, one that effectively corresponds to the programmatic implementation of Shannon entropy.

The authors of AT have advanced the argument that Shannon entropy or LZ compression can only be applied to bits, but this is also mistaken; it can be applied to any object and on any basic units, from vertices in a network to peaks from a mass spectral file, just as they apply their own Ai metric to computer files and computable data representations.

All digital algorithms—including those underpinning AT—operate on representations. Whether it is a mass spectrometry file, a graph of molecular connectivity, or a string representation of a chemical compound, the data is symbolic and abstracted from the physical world. No computer ingests a molecule directly. Not AT, not any other method.

The difference, then, is not in whether abstraction is used, but in how effectively the abstraction captures structure. And here, information theory wins. Compression methods like Shannon entropy or Lempel-Ziv (LZ) not only operate on the same data that AT uses, they do so more efficiently and with mathematically grounded definitions of complexity and redundancy. What AT introduces are additional, ad hoc heuristic layers that obscure rather than clarify the structure being measured.

To make the point more concrete, imagine that a repeating subunit in a molecule, say a benzene ring, is encoded, in a simplified representation, as the binary string 1001. Now consider a polymer such as poly(p-phenylene), composed of multiple such units in a linear chain. Its structure might be symbolically represented as 1001100110011001. A compression algorithm like LZ will quickly identify the repeated pattern and reduce the representation by storing the motif once and indexing its repetitions. From the compressed structure, the original repeating unit 1001 can be fully reconstructed. Rather than broken down, the underlying molecular structure is recovered from the symbolic pattern, even if it has been broken into smaller parts or encoded in a non-chemical alphabet. However, no description of a statistical algorithm includes as a necessary condition that dictionary units or tokens have to be binary code. It can be any alphabet, including the alphabet of molecules that can be forced to not be broken down any further.

The essential insight here is that structure and repetition transcend the encoding format. This is exactly what Shannon formalised in his theory of communication: that the information content of a message is determined by the statistical properties of its components, not by the medium or alphabet used. As long as the representation is lossless and computable, the patterns it contains are physically meaningful and recoverable independent of vocabulary. Turing's universality and algorithmic complexity results extend this principle, showing that any computable transformation, including molecular assembly, can be simulated symbolically with no loss of information.

Yet, one of the inputs used in AT is mass spectrometry data---one of many possible encodings of a molecule's structure. In whatever format it is, it is not the molecule itself, but always a digital representation of it. In this respect, AT does not operate closer to the ``physical'' object than any compression algorithm.

Ultimately, AT's inability to demonstrate superiority over existing measures, combined with its refusal to engage in rigorous comparisons, undermines its validity. Its claims to measure complexity in a novel way remain unsupported, and its insistence on dismissing well-established metrics such as Shannon entropy and LZ compression on the basis that AT explains selection and evolution, despite a lack of any evidence, cannot withstand basic scrutiny.

\subsection{Determinism and Uncertainty}

It may seem paradoxical to think that a theory of Turing machines such as algorithmic probability can act or represent a framework for reasoning under uncertainty, when built entirely on deterministic Turing machines or a universal Turing machine (UTM). Yet, this is precisely what gives it its power. A universal Turing machine, although fully deterministic, can generate all possible computable explanations. The probability assigned to each is then determined by both continuous matching with an unfolding observation and, following algorithmic probability~\cite{solomonoff}, the cumulative weight of the number of short programs that reproduce it, the ultimate accepted Bayesian mathematical theory of induction.

The key insight is that uncertainty is not ontological, but epistemic, just as it is not in the real world (e.g. tossing a coin is fully deterministic, we just model it as a random variable out of our own ignorance and limitations). A Turing machine (TM) itself may not be random, but our ignorance of which specific program may have produced the data introduces the element of uncertainty that is quantified and a confidence value assigned. A UTM that is capable of matching or assigning a new computable model (a TM) to every new piece of information produced, thereby updating its internal belief, functions as a model for scientific induction~\cite{solomonoff}. Rather than modelling stochastic noise or correlations, algorithmic probability ranks explanations in terms of their algorithmic plausibility. It is not about statistical random variables or curve-fitting; it is about causal adequacy and descriptive simplicity. This makes algorithmic probability the optimal solution to the problem of induction, as it formalises the most powerful form of abduction: reasoning to the best explanation. The framework is entirely grounded in the universality of Turing machines, which guarantees that no computable hypothesis space is excluded.

\subsection{On the layered accumulation of conceptual and technical flaws and errors}

The missteps made by the authors of AT stack on one another in a way that is difficult to peel back. The argument advanced to defend the metric not only collapses under its own weight---since, if valid, it would apply equally against their own measure---but also inverts the very role of universality. Universality, whether in computation, construction, or logic, was introduced precisely to dissolve dependence on any particular machine or substrate. To this is added a contrast with Kolmogorov complexity built on a false problem that von Neumann never engaged---the supposed requirement of physical construction---when in fact his universal constructor was only ever a formal existence proof. The outcome is not only technically incorrect and methodologically limited, but also philosophically confused: abstractions are invoked in ways that reverse their original intent.

When framed against the universality of computation and the optimality of algorithmic probability, Assembly Theory is not merely redundant; it is an unnecessary detour that mistakes statistical regularity for causal explanation.

\section{Methods}\label{sectionMethods}

We investigate the counterexamples provided in~\cite{kempes2024} with the purpose of countering what we have identified as fallacious arguments and, misleading and incomplete experiments, including:

\begin{enumerate}[label=(\alph*)]
\item in order for two compression schemes to belong to the LZ family, they must necessarily be identical;
\item in the general case (by accounting for arbitrary strings beyond a handful of pre-selected examples), the expected compression rate (or expected number of factors in their respective parsings) would not be correlated at the asymptotic limit (as the length of strings/objects increases), even though they may not actually be for an a priori chosen set of particular examples (especially for short sequences).
\end{enumerate}

For the Molecular Assembly Index (MA) introduced in~\cite{Marshall2021} as an application of Ai within the context of molecular complexity, we show that it is not only highly correlated with the LZW of InChI strings, but also highly correlated with the length of InChI strings themselves. In addition, we showed that one of the main benefits the authors claim about Ai (`molecular assembly' or MA) as being experimentally measurable using mass spectrometry data, is also observed when applying LZW directly on the InChI chemical compound descriptions, implying that it is neither something of particular interest to report nor something only Ai can do.

In order to expose (a) as false, it is always possible to pick particular cases of strings for which the parsing, according to LZ77, differs from the one according to LZ78~\cite{Cover2005}. These are 2 different LZ compression schemes that differ trivially yet they are based on the same principles, and both are derived and belong to the same family.

One can see in~\cite[Figure~3]{kempes2024} that the authors omit the dictionary built for the right square (i.e., the one for the parsing according to the assembly index calculation method). Specifically, this assembly-coding dictionary is
\begin{equation}
\begin{aligned}
& ( 1 , \textnormal{``z''} ) , \\
& ( 2 , [ 1 , 1 ] ) , \\
& ( 3 , [ 2 , 2 ] ) , \\
& ( 4 , [ 3 , 3 ] )
\end{aligned}
\end{equation}
The final object would be given by the codeword list $[ 4 , 4 ]$ as indicated by the right square in~\cite[Figure~3]{kempes2024}. Notice that the set of vertices of the minimum rooted assembly subspace of the final object is given by the set $\{ 1 , 2 , 3 , 4 , [ 4 , 4 ] \}$ composed of pointers; and the assembly index is the size of this dictionary excluding the basis element under pointer $1$, that is, Ai $= 5 - 1 = 4$. Another fact overlooked by the authors is that if one uses an LZ factorisation/parsing scheme straightforwardly defined by adding to the dictionary the longest prefix of the subsequent subsequence that has already occurred, or a single symbol (in our case, ``z'') if there is no such nonempty prefix, then one would obtain exactly the desired parsing
\begin{equation}\label{eqTrivialparsing}
\textnormal{z} |
\textnormal{z} |
\textnormal{zz} |
\textnormal{zzzz} |
\textnormal{zzzzzzzz}
\end{equation}
that the authors of AT overlook, a parsing which gives the exact assembling process in the minimum rooted assembly subspace of the final object in Equation~\eqref{eqTrivialparsing}. Notice that the number of factors/codewords in Equation~\eqref{eqTrivialparsing} (minus 1) is the assembly index of the final object. Except for the initial basis element ``z'', the parsing in Equation~\eqref{eqTrivialparsing} exactly corresponds to the sequence of edge labels of the rooted assembly path of the final object. In this trivial example, we see that the shortest rooted assembly path and the longest one (which belong to the minimum rooted assembly subspace) are the same.

This also answers the AT authors' challenge and questioning in~\cite{kempes2024} regarding the lack of dictionary or memory in Ai, this is yet another fallacy, the dictionary is in the parsed molecules themselves. It is not that they do not. and that LZ algorithms are artificial computational constructs, it is just that the authors have not been capable of finding the obvious technical analogies and equivalences that in computer science are separated for implementation or pedagogical purposes.

Regarding (b), the authors' counterargument constitutes a combination of the ``Texas sharp shooter'' and the ``hasty generalisation'' fallacies: where one takes the global properties as always the same as the properties resulting from analyses of particular local cases, which they would invariably be, given that the criteria for the latter analyses were chosen to produce the desired local properties.

From the equivalence between Ai and dictionary-based compression schemes formally demonstrated in~\cite{abrahao2024}, we know that the set of vertices/objects of the minimum subspace encodes/compresses the final object (in our case of interest, the input string). This holds whether the basis elements are included or not, since the compression rate obtained in the former case or the latter case differs only by an independent constant. Thus, the number of codewords in the list (which results from the parsing of the string according to the assembly index method) is equivalent to the length of the longest (rooted) assembling paths in the minimum subspace, except for those that are basis elements. In other words, they show all the elements of a typical statistical compression algorithm as proved to be.

For example, for the input string ``abracadabra'', the dictionary derived overlooked by the authors of AT in their example calculations is:
\begin{equation}
\begin{aligned}
\textnormal{`b'} &: \textnormal{`!'}, \\
\textnormal{`r'} &: \textnormal{`@'}, \\
\textnormal{`a'} &: \textnormal{`\$'}, \\
\textnormal{`c'} &: \textnormal{`\%'}, \\
\textnormal{`d'} &: \textnormal{`\&'}, \\
\textnormal{`ab'} &: \textnormal{`('}, \\
\textnormal{`abr'} &: \textnormal{`)'}, \\
\textnormal{`abra'} &: \textnormal{`*'}
\end{aligned}
\end{equation}
while the codeword list is
\begin{equation}
\textnormal{[[`\$', [`!'], [`@'], [`\$']], [`\%'], [`\$'], [`\&'], [`*']]. }
\end{equation}
Thus, Ai is simply $8 - 1 = 7$, which is given by the number of single symbols inside the operator $[\cdot]$ that represent an edge in a rooted assembling path whose label is taken from one of the vertices of the minimum subspace.

For LZW, the dictionary would be
\begin{equation}
\begin{aligned}
\textnormal{`b'} &: \textnormal{`000'}, \\
\textnormal{`r'} &: \textnormal{`001'}, \\
\textnormal{`a'} &: \textnormal{`010'}, \\
\textnormal{`c'} &: \textnormal{`011'}, \\
\textnormal{`d'} &: \textnormal{`100'}, \\
\textnormal{`ab'} &: \textnormal{`101'}, \\
\textnormal{`br'} &: \textnormal{`110'}, \\
\textnormal{`ra'} &: \textnormal{`111'}, \\
\textnormal{`ac'} &: \textnormal{`0000'}, \\
\textnormal{`ca'} &: \textnormal{`0001'}, \\
\textnormal{`ad'} &: \textnormal{`0010'}, \\
\textnormal{`da'} &: \textnormal{`0011'}, \\
\textnormal{`abr'} &: \textnormal{`0100'}
\end{aligned}
\end{equation}
while the codeword list is
\begin{equation}
\textnormal{[ `010', `000', `001', `010', `011', `010', `100', `101', `111' ].}
\end{equation}
Thus, LZW = 9.

In the case of LZW/LZ78 we know that the size in bits of the compressed form of the input string (in the asymptotic limit) grows under $c_1(n) \mathbf{O}(\log(c_1(n)))$~\cite{Cover2005,Salomon2010HandbookDataCompression}, where $c_1(n)$ is the size of the LZW/LZ78 dictionary for a string of length $n$. The same also holds for the compressed form (with size $K_{Ai}(x)$ in bits) of the input string $x$ according to the assembly index calculation method, so that
\begin{equation}
K_{Ai}(x) \leq c_2(n) \mathbf{O}(\log(c_3(n))) \leq c_2(n) \mathbf{O}(\log(c_2(n))) \text{ ,}
\end{equation}
where $c_2(n)$ is the length of the list of codewords that constitutes the string $x$ and $c_3(n) = \mathrm{Ai}(n)$ is the assembly index that is the size of the Ai dictionary~\cite{abrahao2024} for a string $x$ of length $n = |x|$ with $\mathrm{Ai}(n) \leq c_2(n)$. This shows that Ai does implement an implicit dictionary in accordance with the expectation of belonging to the family of popular statistical compression algorithms LZ that is widely used in complexity science.

Sequences with larger values of $c_i(\cdot)$, i.e., the number of distinct phrases or codewords in the dictionary, to which the pointers necessarily recur less often, would have a lower probability. This is the same key idea used to demonstrate the optimality of LZ schemes toward the limit given by the noiseless coding theorem; and it is the same idea behind the assembly index as a complexity measure, which the authors claim is a novelty.

When it comes to molecules, the dictionaries are simply embedded in the molecules themselves. Every time their representations are parsed and counted, a dictionary is actually being built behind the operation of the Ai metric.

Here we further investigate the relationships between these values with experiments that generalise the purported counterexamples to much longer sequences of different configurations, as explained in Section~\ref{sectionAiconvergences}, thus demonstrating that (b) is indeed such a fallacy.

\subsection{Stochasticity and Connections of BDM with Shannon Entropy and LZW}\label{sectionBDM}

In its expanded form, the BDM of an object $x$, according to the partition $i$, is defined by
\begin{equation}\label{eqBDM}
BDM( x , i , m ) = \sum_{(r_j , n_j) \in P_i(x)} (\log(n_j) + K_m(r_j)) \text{ ,}
\end{equation}
\noindent where:
\begin{itemize}
\item $P_i(x)$ is the set of pairs $(r_j, n_j)$ obtained when decomposing the object $x$ according to a partition $i$ of contiguous parts $r_j$, each of which appears $n_j$ times in such a partition (i.e., $n_j$ is the multiplicity of $r_j$).
\item $K_m(r_j)$ is an approximation to $\mathbf{K}(r_j)$ and $m$ is the index of approximation method employed to calculate $K_m(r_j)$.
\end{itemize}

Notice that Equation~\eqref{eqBDM} can be expressed for many particular cases, e.g. in the case of unidimensional objects (such as bitstrings) but also any other object where $m$ indicates the method given by the pre-computed table of complexity estimations derived from CTM~\cite{zenilbdm,iscience,zenil2018review,zenilreview}.

Also notice that if one takes a partition $i_0$ for which the object's size is the partition size, then $BDM(x, i_0, m)$ becomes equivalent to the approximation method indexed as $m$, e.g. the CTM. That is, there is $m$ such that
\[ BDM(x, i_0, m) = CTM(x, Q) \text{ ,} \]
where $Q$ is a computable (proper) subset of $\{\mathbf{M} : \exists w' (\mathbf{M}(\emptyset) = w')\}$ and $Q(x) \subset \{\mathbf{M} : \mathbf{M}(\emptyset) = x\}$ a computable subset of $Q$ that the method $m$ calculates, where the algorithmic probability (AP) of $x$ is approximated by
\begin{equation}\label{eqCTM}
\frac{Q(x)}{Q}
\end{equation}
and $\mathbf{M}(x)=y$ denotes algorithm $\mathbf{M}$ returning $y$ with input $x$.

Deriving from this equivalence, one of the properties of BDM is that it is a complexity measure that approaches $\mathbf{K}(\cdot)$ in the best case scenario, and approaches block (Shannon) Entropy in the worst case~\cite{zenilbdm}, that is,
\begin{equation}\label{eqBDMbounds}
\mathbf{K}(x) \leq \mathbf{O}(1) + BDM(x, i_0, m) \leq \mathbf{O}(1) + BDM(x, i, m') \leq \mathbf{O}(1) + N_i \mathbf{H}_i(X^i) \text{ .}
\end{equation}

The left inequalities of Equation~\eqref{eqBDMbounds} always hold, from the basic properties of algorithmic complexity. From classical information theory, the rightmost inequality holds for sufficiently large $|x|$ and any computable compression method $m'$ that is capable of matching (or surpassing) the statistical compression/encoding achieved by, e.g., Huffman or LZW for each codeword, where
\begin{equation}
\mathbf{H}_i(X^i) = - \sum_{(r_j, n_j) \in P_i(x)} p(r_j) \log(p(r_j))
\end{equation}
is the \emph{block} (Shannon) Entropy $\mathbf{H}_i$ of an i.i.d. source $X^i$ such that $p(r_j) \to \frac{n_j}{N_i}$ as $|x| \to \infty$, the random variable $X^i$ can assume values in the set $\{r_1, \dots, r_j, \dots, r_{|P_i(x)|}\}$, and $N_i = \sum_{(r_j, n_j) \in P_i(x)} n_j$.

Thus, in case $m'$ is a sufficiently powerful computable approximation and $i$ a partition closer to the partition $i_0$, one obtains $\mathbf{K}(x) \sim BDM(x, i, m')$; otherwise, in case the object $x$ is indeed generated by a stochastic source, one obtains $BDM(x, i, m') \leq \mathbf{O}(1) + N_i \mathbf{H}_i(X^i)$ approaching the equality the more statistically random the source is. Also note that, as demonstrated in~\cite{zenilbdm} for regular bidimensional partitions, an algorithm can always losslessly retrieve $x$ from the information in the parts/blocks calculated in $BDM(x, i, m')$ by an encoding scheme $\left< \cdot \right>_{BDM}$ such that
\begin{equation}
|\left< x \right>_{BDM}| \leq \mathbf{O}(1) + BDM(x, i, m') + \mathbf{O}(N_i \log(|P_i(x)|))
\end{equation}
holds in bits.

This means that for objects generated according to a probability distribution sufficiently close to the uniform distribution and a sufficiently powerful $m'$, BDM can always achieve a statistical lossless compression of the same order as $\mathbf{H}$, while the value of $BDM(x, i, m')$ can be smaller than $N_i \mathbf{H}_i(X^i)$. Therefore, $\mathbf{H}$ is a (statistical) compression method subsumed within BDM.

A false argument repeated by the authors of Assembly Theory is that approximations to algorithmic Kolmogorov complexity are uncomputable, unlike assembly theory~\cite{Marshall2022TheoreticalAT}. Not only is this false, as BDM is fully computable~\cite{soler2017computable,soler2013correspondence}, but also, the assembly index itself is an approximation to $\mathbf{K}$ equivalent to $\mathbf{H}$ and LZ compression, as we have proven~\cite{abrahao2024}, therefore a loose upper bound on $\mathbf{K}$ itself. Hence, the assembly index offers no advantage.

BDM can also deal with stochasticity, by exploring all possible computer program pathways that assemble an object, as fully explained in Section~\ref{sectionPhysicalGrounding}.

\subsection{Sequence Convergence}\label{sectionAiconvergences}

In the one example provided in~\cite{kempes2024}, the authors quantify the correlation of their assembly index and LZW, showing a low statistic and claiming it as a proof of dissimilarity vis-a-vis LZW. The string used is very short and of fixed length. It's known that LZW produces codewords that affect shorter strings, as it carries the information to decompress a string while the assembly index discards it. Here, we repeat their experiment on larger and more diverse strings, showing a full correlation with LZW and with Shannon entropy.

In our experimental case, we took three types of strings:
\begin{itemize}
\item \textbf{ZBC sequence:} zbzbczbzbczbzbc, as used in~\cite{kempes2024} to prove the dissimilarities between the assembly index and LZW.
\item \textbf{Patterned sequence:} 2 repeated patterned sequences of fixed length, the first ABCDE and the other ABCDEFGHIJ, repeating to a length of 100.
\item \textbf{Random sequence:} Random sequences, generated as 2 random strings of up to length 100. The growing permutations of the ZBC sequence also fall within the category of random sequences.
\end{itemize}

For growing sequences in each case, we simply add more ZBC morpheme blocks as suffixes or add more random characters.

\subsubsection{Fixed vs. Varying Lengths of ZBC Strings}

10,000 random permutations were generated for the ZBC sequence in both fixed and varying random lengths, followed by complexity measure analyses. The ZBC sequence underwent random permutation to generate 10,000 variants for each length category. In short, to generate the strings the algorithm basically picked $n$ characters 10,000 times from the string zbzbczbzbczbzbc. In our case, $n$ ranged from 2 to 100 for some plots and from 2 to 200 for an in-depth analysis of the long-term convergence between LZW and Ai.

The complexity measures computed included Lempel-Ziv-Welch complexity (LZW), Shannon entropy Rate, and Assembly Index. LZW and assembly index (Ai) calculations were previously described in Section~\ref{sectionMethods}. Shannon entropy and its rate were calculated to quantify the statistical randomness of the sequence, while BDM evaluated the statistical and algorithmic complexity of the permutations. Density plots and statistics are provided to quantify the relationship between LZW and the Assembly Index for both fixed and varying lengths.

For sequences obtained by considering strings with lengths ranging from 2 to 200, we studied the asymptotic behaviour of LZW and Ai as well as of the ratio between the logarithms of these metrics. We studied how the mean and standard deviation of this ratio change as the length of the strings grows. We fitted power laws to these two statistics to extrapolate the behaviour of larger strings.

\subsubsection{Growing Sequence Analysis}

Lastly, we analysed linear chains (sequences) growing from lengths 1 to 3000. Two sets of sequences were studied: random and patterned sequences (patterns of repeated length 5, and 10). The patterns of length 5 consisted of repeated blocks of the first 5 alphabets ABCDE, and similarly for the repeated block patterns of 10. For each length increment, the complexity measures were computed. Correlations between the Assembly Index and other complexity measures were calculated every few steps in the growth process. Density plots and growth measure plots for each sequence were generated. This provides a quantitative analysis of the complexity dynamics as sequences increased in length.

Pearson and Spearman correlations between the Assembly Index and the other complexity measures were calculated at intervals of certain steps (permutations) to identify significant relationships, providing insights into the structure and randomness or patterns across the length and sequence types of each growing string. A Pareto (long-tail) distribution fit was assessed on all BDM distributions, while a Gaussian curve fit was assessed on the LZW-assembly index density plots. All other distributions for other complexity measures underwent a Bernoulli curve fit.

As shown in Figures~\ref{main}C and~\ref{main}D, a near-perfect correlation between LZW and the Assembly Index is observed across all conditions (patterned and random) growing sequences. Table~\ref{tabpatterned} displays the averaged Spearman correlations between the assembly index (Ai) and other complexity measures (LZW, Entropy Rate, BDM) for patterned sequences of increasing length. LZW and BDM measures are the most robust, consistently converging to a correlation of 1 across all sequence lengths. The average correlation between LZW and Ai was 0.98-0.99. Similarly, Table~\ref{tabrandom} displays the convergence of all three complexity measures: LZW, entropy rate, and BDM to an average Spearman correlation of 1 for random growing sequences shown in Figures~\ref{main}A and~\ref{main}B.

\section{Conclusions}

\begin{table}[h!]
\centering
\caption{What we have demonstrated based on mathematical and empirical evidence.}
\label{table1}
\scriptsize
\setlength{\tabcolsep}{3pt}
\renewcommand{\arraystretch}{1.05}
\arrayrulecolor[gray]{0.8}

\begin{tabularx}{\textwidth}{
  >{\raggedright\arraybackslash}p{0.24\textwidth}
  >{\raggedright\arraybackslash}p{0.4\textwidth}
  >{\raggedright\arraybackslash}p{0.35\textwidth}}
\toprule
\textbf{Claim by AT/Ai} & \textbf{Our counter-argument} & \textbf{Technical basis} \\
\midrule
AT/Ai explain selection/evolution & 
Ai measures redundancy, not environment-dependent selection & 
Selection depends on $(o,e)$; Ai $\sim$ LZ cannot encode fitness or counterfactuals \\
\midrule
AT unifies physics and biology & 
No new invariant/law; Ai reduces to entropy-like redundancy & 
Representation-dependent score; no conserved quantity or predictive law \\
\midrule
Ai is not data compression & 
Ai is a dictionary parsing (statistical compression) algorithm proven mathematically with an undisputed theorem~\cite{abrahao2024} & 
Equivalent to LZ-style factorisation; code-length $\propto$ Ai output~\cite{Uthamacumaran2024,abrahao2024} \\
\midrule
Ai is unrelated to Shannon or Kolmogorov complexity & 
Ai approximates expected code length; loosely upper-bounds $K(x)$ & 
Hierarchy: Ai $\equiv$ LZ $\preceq H \prec$ BDM $\preceq$ CTM $\prec$ AP $\equiv K$ \\
\midrule
Kolmogorov complexity is unusable because it is uncomputable & 
$K$ is semi-computable; many branches of mathematics rely on uncomputable concepts (e.g., calculus, analysis) & 
BDM/CTM give computable $K$-proxies with many applications~\cite{springerbook,cupbook} \\
\midrule
AT avoids Turing-machine limits & 
Ai is computed on a computer program, hence a Turing machine; appealing to time classes is inconsistent & 
Rejecting TMs while invoking time complexity is self-contradictory~\cite{kempes2024} \\
\midrule
Compression destroys physical meaning & 
Compression units or tokens can represent molecules rather than binary strings; the objection confuses algorithm with dictionary particularities & 
Fixing dictionary units to molecular granularity keeps objects intact but even `destroying' them, they are recovered from their patterns \\
\midrule
Ai uniquely captures stochastic/causal structure & 
Algorithmic probability and Shannon entropy already handle uncertainty; Ai adds extra steps and unfounded claims & 
Shannon entropy, Kolmogorov complexity, and algorithmic probability subsume copy-counting~\cite{abrahao2024} \\
\midrule
Any positive result reported by AT is not the consequence of a unique breakthrough or feature of AT but the direct result of computing Shannon entropy under another name & 
The same outcomes, including cutoff values distinguishing organic from inorganic compounds, have been previously reported in the literature using entropy and compression-based analyses, as well as methods that go beyond counting trivial repetitions and justify causality~\cite{zenil2018} & 
AT reproduces prior entropy-derived results and presents them as novel while offering no new theoretical or empirical contribution~\cite{zenil2018,hernandez2018} \\
\bottomrule
\end{tabularx}
\end{table}

\begin{table}[h!]
\centering
\caption{What we have neither claimed nor advocated for.}
\label{table2}
\scriptsize
\setlength{\tabcolsep}{3pt}
\renewcommand{\arraystretch}{1.05}
\arrayrulecolor[gray]{0.8}

\begin{tabularx}{\textwidth}{
  >{\raggedright\arraybackslash}p{0.30\textwidth}
  >{\raggedright\arraybackslash}p{0.68\textwidth}}
\toprule
\textbf{Not our claim} & \textbf{Our position / clarification} \\
\midrule
Statistical compressors should replace AT as a theory of selection and evolution & 
Compression serves as a null benchmark revealing Ai's redundancy, not a theory of life. AT and Ai implement statistical compression and is as inadequate to characterise evolution and selection as any other statistical compression algorithm\\
\midrule
Ai is inefficient and should be swapped for faster compressors & 
Our critique is conceptual and foundational; runtime is not the issue. It is the AT authors who claim that because their algorithm is slower than statistical compressors, it is different and better with no evidence\\
\midrule
Molecules are just strings & 
The application of data compression to e.g. SMILES molecular descriptions in our experiments had the only purpose to serve as a control experiment to show that it produces the same results that AT claims is capable of with Ai on experimental data \\
\midrule
Shannon entropy or Kolmogorov complexity can explain selection or evolution & 
We do not reductionistically derive biology from either measure; we show that AT adds no new insight, whereas Kolmogorov complexity can in fact capture causality unlike AT \\
\midrule
Entropy should replace AT as a theory & 
We advocate removing methodological redundancy, not elevating $H$ to a theory of life \\
\midrule
Algorithmic similarity implies ontological identity & 
Informational equivalence shows lack of novelty, not sameness of phenomena \\
\bottomrule
\end{tabularx}
\end{table}

Given that our position has been oversimplified by some authors, who have implied that we advocate the use of compression algorithms or their application to SMILES as an alternative to Assembly Theory (AT)~\cite{newpaperoyal}, we have clarified and summarised our position in Tables \ref{table1} and \ref{table2}. In fact, our argument is the opposite: we have shown that AT, when implemented with Ai, is effectively a trivial statistical compression algorithm. As such, it cannot account for selection or evolutionary processes any better than other algorithms in the same class. Moreover, applying any statistical compression algorithm to SMILES molecular representations reproduces the same results reported by AT, thereby casting doubt on its novelty or the fundamental nature of its contribution.

As we have shown, it is not possible for Assembly Theory (AT) and its Assembly Index (Ai) to explain or quantify selection or evolution because of its trivial nature, which does not distinguish it from Shannon entropy, neither fundamentally nor judging by its results. Instead, AT's sophisticated jargon collapses the distinction between causal and descriptive domains, confusing the historical processes that generate diversity with the static configurations that these processes leave behind. Contrary to their claims, AT takes objects as instant and final, for which they produce plausible statistical, not causal, origins with no evidence other than chemical or molecular constructibility completely disconnected from evolution and selection as Ai only captures the minimal number of trivial combinatorial steps required to build a molecule, rather than the causal or heritable dynamics through which biological change unfolds. This makes AT fundamentally incapable of representing evolutionary mechanisms such as variation, inheritance, and selection, which depend on temporally ordered transformations rather than structural enumeration.

The assumption that statistical repetition (molecular or otherwise) mirrors evolutionary progression fails in reducing all differences to a superficial similarity metric. In essence, AT substitutes pattern resemblance for causal explanation, offering a phenomenological index of statistical difficulty that Shannon entropy already captures but without accounting for how or why those statistical regularities may accumulate over time. AT confounds speculation with theory. 

This paper completes the theoretical and empirical demonstration that any variation of the copy-number concept underlying AT and Ai, which involves counting the number of repetitions in an object, such as a molecule, to arrive at a measure of selection or evolution, is equivalent to popular statistical compression algorithms based on Shannon entropy. We showed that the authors' argument that the intractability of AT separates Ai from standard compression does not withstand basic scrutiny and is not robust, and that their empirical results separating organic from inorganic compounds, on which their whole empirical evidence is based, have not only been previously reported, avoiding claims of grand unification between physics and biology, but are mostly driven by molecular length---which the authors did not control for.

Given that Ai can only account for basic statistical operations that cannot account for the full extent and complexity of selection and evolution, this contribution makes AT unsuitable and redundant, especially when applied to their own experimental data, on which we have proven that both Shannon entropy and LZ compression reproduce the same or even better results without the extra steps and the additional unjustified assembly argot that sometimes can be useful if honest, open and transparent.

What AT has offered is an obscure theory-of-everything narrative with little substance and a refusal to engage with control comparisons or foundational mathematics. At best, we consider Assembly Theory (AT) a pedagogical exercise, potentially useful for illustrating certain patterns of structural repetition in molecular complexity that might or might not lead to future insights. However, this limited didactic value stands in clear contrast to the sweeping claims made by its proponents in high-profile media coverage, exaggerated press releases, and academic publications. AT does not explain evolution or selection; it does not unify physics and biology; and it does not redefine complexity. As our analyses demonstrate, AT simply reframes familiar statistical compression methods in unfamiliar terminology, introduces unnecessary computational steps, and presents the result as if it were a scientific breakthrough.

\section{Funding}

Felipe S. Abrah\~{a}o acknowledges support from the S\~{a}o Paulo Research Foundation (FAPESP), grants $2021$/$14501$-$8$ and $2023$/$05593$-$1$. No other funding was received.

\newpage

\bibliographystyle{sn-nature}
\bibliography{biblio}

\newpage

\noindent \\
\begin{appendices}

\renewcommand{\thefigure}{S\arabic{figure}}
\renewcommand{\thetable}{S\arabic{table}}
\setcounter{figure}{0}
\setcounter{table}{0}

\section*{Supplementary Information}

\subsection*{More Context and Literature on Chemical and Molecular Complexity}

Assembly theory has undermined long-held ideas in algorithmic information theory, but also elements of influential work throughout complexity theory and systems science, such as statistical theories of molecular evolution~\cite{kimura1974principles}, the nonequilibrium thermodynamics of self-organisation~\cite{schrodinger1944life,prigogine1980being}, and evolutionary dynamics conceptualised as multi-level teleonomic processes or ecosystem models of multiscale interactions\cite{corning2023evolution}. 
For instance, Nobel Laureate and complex systems pioneer Herbert Simon argues that complex systems evolve hierarchical and multi-nested patterns or networks through stable intermediate forms, allowing rapid emergence through scaling of emergent mesoscale structures at distinct levels, from simple to complex~\cite{simon1962architecture}. 
According to physicist Freeman Dyson, for early life to arise and evolve, it needed a means of storing and replicating information with a low error rate~\cite{dyson2008origins}. This aligns with ideas in algorithmic information theory about the need for mechanisms to compress information and replicate it accurately. Hence, the ideas of systems theorists such as Simon, Dyson, Prigogine, and Schr\"{o}dinger about complex systems and negentropy-based self-organisation in living systems, which are essentially saying that life can reduce entropy/disorder in local regions, and that these ordered structures must be able to maintain and transmit their organisation in the form of redundancies. 

For example, in 2005, Cilibrasi and Vitanyi showed that by finding redundancies in data, such as identical copies, they could fully reconstruct an evolutionary genetic tree~\cite{cilibrasi}. These ideas have been formalised and empirically supported within the framework of algorithmic information theory and its fundamental notions and computable methods of compression. 

Life can describe itself in a more compact way algorithmically through processes like genetic programs with low Entropy, low algorithmic complexity/high compressibility~\cite{zenilgershenson,iscience}. That compression or counting copies works does not come as a surprise. In molecular biology, for example, it is well established that counting for GC nucleotides, which are molecules, discloses the relationship between species. This is because two species that are evolutionarily related to each other will have about the same number of GC nucleotide copies. This is referred to as GC content. We have conducted research showing how information theory and algorithmic complexity can be combined with and compared to GC content to discover DNA transcription regions in nucleosomes of high genomic content~\cite{zenilnar}.

Biophysicist Harold Morowitz discusses how the core biochemical pathways of life, such as the reductive citric acid cycle and biosynthesis of complex biomolecules, can be understood as order parameters in a phase transition driven by free energy flows~\cite{morowitz2007}. The spontaneous emergence and stabilisation of such ordered biochemical networks are explained using ideas from nonequilibrium statistical mechanics, where maximising the structure's chemical network entropy accounts for the flows of matter and energy through reaction channels, allowing for self-organised criticality and dissipative structures driven by environmental free energy~\cite{morowitz2007}.
Morowitz proposed that the bioenergetics of living systems self-organise energy dissipating pathways that assemble into chemical structures~\cite{morowitz1978foundations}. Chemical network entropy or graph entropy explains how this buildup of energy in the chemical system increases entropy and drives the self-organisation of assembly pathways such that the emergent structures maximise the flow of energy through the chemical network~\cite{morowitz1978foundations}. Schrodinger coined the term negentropy to describe such processes of self-assembly. Many others have contributed to similar statistical thermodynamics of biomolecular evolution using information-theoretic concepts of dissipative energy and chemical network entropy in open systems far from equilibrium~\cite{prigogine1980being,vanchurin2022}. Information theoretics of complex networks explain life's evolvability and the molecular evolution of chemical networks. Even as early as the 1950s, pioneers like Rashewsky applied entropy formulas to quantify the structural information content of biological networks~\cite{dehmer2008}.

For instance, Nowak and Ohtsuki~\cite{nowak2008} formulated a theory of pre-evolutionary dynamics where random polymerisation of monomers leads to a generative system with selection and mutation before replication. The branching tree structure of combinatorial replication in this `prelife' system can be analysed using compression schemes like Huffman coding from algorithmic information theory, and concepts like entropy, to quantify the information flow and diversity produced during the emergence of evolutionary dynamics~\cite{nowak2008}. England discusses how self-replicating systems must generate entropy to fuel their replication, even when far from equilibrium~\cite{england2013}. He posits relationships between a replicator's maximum growth rate, its internal entropy production, robustness, and the amount of heat dissipated. This echoes autopoietic theories of life proposed by Maturana and Varela, which define living beings as self-generating networks that produce their own components and sustain their self-organisation through ongoing interactions with their environment~\cite{maturana1980,england2013}. Both England and autopoiesis view self-reproduction in thermodynamic terms, requiring a constant throughput of energy and matter to maintain a system's complex internal organisation.

Equipped with all this knowledge and based on the principles of classical information theory and algorithmic information theory, in a series of papers spanning the first decade of the 2010s we covered and elaborated on every idea that the authors of Assembly Theory claim to have introduced for the first time~\cite{zenilacm,soler2017computable, soler2013correspondence,zenilcausal}, including in an article featured in the last referenced publication,
a book edited by one of the senior authors of Assembly Theory (AT), an article from which the authors (the other senior author having also published in the same book) seems to have replicated our ideas. In particular, we introduced a measure called the Block Decomposition Method (BDM)~\cite{zenilbdm} that as a baseline, counts identical copies in data, but also many other variations that form causal modules~\cite{zenilctm} in connection to selection and evolution~\cite{hernandez2018}.

In the early 2010s, a quantitative measure of physical complexity based on the information required for self-assembly, applicable to any structure, including chemical and molecular networks, was introduced in connection to evolution~\cite{anhert}. By quantifying the complexity of molecules and protein complexes, the approach highlighted how symmetry and modularity are favoured in biological self-assembly, and offered metrics for comparing different physical structures. With the same group, we also reported that a measure of algorithmic complexity of graphs correlated equivalent networks, revealing a connection between automorphism size and algorithmic information content, and showing how to classify real-world graphs, including metabolic networks~\cite{ardzenil} as in~\cite{zenilbibm}. In this other work in the mid 2010s, we also showed how approximations of algorithmic complexity could quantify the self-assembly properties of nonDNA molecular computing in application to an example of synthetic porphyrin molecules (the molecules that give their colour to red blood cells)~\cite{kraszenil}.

Then, in 2018, almost four years before Assembly Theory's 2021 paper on the topic~\cite{Marshall2021}, and based on our previous work and indexes, we showed how to characterise chemical compounds and structures by using approximations of classical and algorithmic information theory with complexity measures like entropy and BDM~\cite{zenil2018}, which Assembly Theory would rediscover and sell as a unification of biology and physics~\cite{Sharma2023}. 
Graph entropy and similar algorithmic information theoretic measures capable of quantifying network topology~\cite{zenil2018review} can help explain molecular chemical evolution by characterising how structural information is processed and flows within biochemical~\cite{zenil2018} and genetic networks~\cite{iscience}. 
The principles of algorithmic probability can be applied to make quantitative predictions on chemical structure networks to profile their causal properties, track changes, and investigate the relationship between their topological or physical properties and the algorithmic sensitivity of chemical constituents, such as elements and bonds, during chemical processes~\cite{zenil2018}.

Other systems science and complexity theorists have also demonstrated that algorithmic complexity is a more robust measure than graph entropy for quantifying a system's complexity and evolvability~\cite{zenil2018review,morzy2017measuring}. Such concepts can explain the emergent behaviours of life, even at higher orders and multiscalar processes. For instance, the insights of evolutionary graph theory into higher-order network motif impact on fixation times have implications for designing populations to optimise the discovery of solutions in complex adaptive landscapes, pointing to teleonomic processes at work in molecular evolution networks~\cite{kuo2024}. 
Evolutionary game theory-based models using concepts of evolutionary strategies for the optimisation of search space in molecular constructions, and replicator equations, can serve as alternatives to chemical graph/network entropy for explaining the stability and emergence of complex molecular networks~\cite{Veloz2014, Cressman2021}.
Based on algorithmic information theory, but also at the intersection of other fields like distributed computation, complex networks, and cybernetics, it has been demonstrated that both game-theoretical approaches and network complexity can be unified into a theoretical framework~\cite{Abrahao2017publishednat,Abrahao2018ComplexSystemsjournal,Abrahao2021bEmergenceAIDPTRSA} that subsumes both evolutionary processes and network dynamics, e.g. for both local selection, adaptative and global synergistic relationships.

More recently, process biologist Jaeger has pointed out that Assembly Theory does not explain or quantify evolution and selection as claimed, and uses these terms too broadly. Jaeger notes that Assembly Theory's use of metrics like a structure's self-assembly through recursively combining objects over minimal paths and predicted abundance based on rule weights is not novel, as they are ``special cases'' of previously established metrics~\cite{jaeger2024}.

Due to the multiscalar complexity and goal-directed behaviours exhibited across living and engineered systems, the traditional binary categories of organic and inorganic do not fully capture this continuum or evolutionary (adaptive) spectrum. Emerging fields in synthetic biology and soft matter systems are demonstrating that life comprises diverse, hybrid forms with programmable, self-producing (autopoietic) properties.
Further discussion of extrinsic vs. intrinsic complexity measures can also be found in~\cite{abrahao2024}.

While the strategy of calling Shannon entropy and LZ compression by another name has proven fruitful for the authors of Assembly Theory, it has done a disservice to science and good scientific practice and communication. It has proven to be very successful in breaking into the review and news cycle systems, conferring on it a sense of breakthrough and novelty. However, all of AT's results can be reproduced with naive statistical metrics without resorting to new theories or indexes that make grandiose claims, as we have shown here and in previous work~\cite{Uthamacumaran2024,abrahao2024}.

The self-assembly of inert and living matter has been widely recognised and understood for centuries, with many assembly processes explained by physical laws and fundamental principles---from galaxies forming due to gravity to animals and organs constructed from a small subset of polymers. Although ignorance has sometimes helped spur new discoveries, Assembly Theory does not offer new insights into the mechanisms driving these assembly processes and is not more capable of quantification than other algorithms in the existing literature, some of which have been previously reported but have been overlooked by the authors~\cite{zenil2018,hernandez2018}.

\subsection*{Misleading Arguments about Compression, Uncomputability, and Complexity measures}

On the overreaction against compression: the scientific endeavour is about compression through model simplification, to explain more with less, effectively compressing observed data into scientific theories. In fact, compression has been shown to be equivalent to prediction~\cite{Downey2010}, and even equivalent to comprehension, broadly speaking~\cite{zenil2020CompressionComprehension}, so it is a fundamental concept of science~\cite{kirchherr1997Miraculous}, not just computer science, and not a technicality bearing no relevance to nature. 
It is a universal assumption in the practice of science based on Ockham's razor~\cite{smalldata}: the better the compression, the more likely a model or theory is to capture a natural phenomenon and the more accurately it may predict its behaviour. This has been a dictum of science since the times of Galileo, is foundational to formal science, and is at the core of Western philosophy, and it has nothing to do with the technicalities of compression, which the authors of AT seem to overreact to in their attempt to distance themselves from actual specific algorithms introduced in the 1970s~\cite{lzw}, from which theirs happen to be indistinguishable.

The authors of AT appear not to realise the value of compression as a fundamental concept and practice in science, and how explanation and model are deeply connected to compression~\cite{bibid} both statistically and algorithmically. Compression and prediction have been formally connected~\cite{chaitin,kolmo1,kolmo2,solomonoff} showing that if one can compress an object, one has gained an understanding of the underlying patterns that allow prediction; if not, then prediction is impossible~\cite{Downey2010}. 
These connections were made by the founders of algorithmic information theory, showing an equivalence between compression, prediction and universal statistical tests for randomness, assuming nothing else but the concept of computation and the existence of computable processes.



The sort of crude understanding of complexity hierarchies and the relationship with other indexes displayed in e.g.~\cite{kempes2024,Marshall2022TheoreticalAT} explains some of the field's shady reputation at times. Authors keep producing indexes that simply turn out to be redundant, without a deep understating of the way in which they can be inserted into the complexity landscape, and by neglecting to perform the most basic control experiments and comparisons, new metrics can make a disservice to the field, especially when introduced to the public together with unfounded claims, such as the claims that AT unifies biology and physics by way of simply counting repetitions in data.


If Assembly Theory is looking for the most likely causal explanation for agnostic selective and evolutionary choices (i.e., without taking into consideration the agent's interaction with its environment), then it is not copy number but algorithmic probability~\cite{nmi2019,iscience} that AT should be studying and applying~\cite{abrahao2024}.

Under Algorithmic Information Theory, algorithmic probability is the other side of the algorithmic complexity coin, where the measure is no longer defined on the shortest program that produces an output, but on the set of all the causal and mechanistic explanations that account for an observation shorter than the description of the observation itself. 

AT has (re-)introduced this notion as if it were its own original contribution, albeit based on a watered-down version of algorithmic probability. In other words, it is currently claiming all the attributes of algorithmic probability and algorithmic complexity while professing to be only about copy number, and implementing another variation of Shannon entropy-based complexity measures, and therefore a much more restrictive (proper) subset of what algorithmic complexity can measure.

It effectively implements weak approximations to these two mainstream measures, but without disclosing this fact;
it eschews any claim to be like or to share the principles and goals of algorithmic information theory, while in fact implementing Shannon entropy again, acknowledging neither of them and representing both as new and an original contribution of AT.

In practice, all real-world applications of Kolmogorov complexity involve computable approximations, such as LZW, CTM, or BDM, each progressively approaching the uncomputable ideal. Ai, being computable and empirically defined, belongs squarely among these approximations rather than transcending them.

Our findings demonstrate that the authors of Assembly Theory still have to explain, account for, and show how their assembly index is different from---and more useful to--- existing indexes, in particular those of the LZ family (and, as demonstrated in this article, even LZW when used on more than a narrow selection of examples). There has been a traditional complexity science or complex systems community absorbed in producing endless variations of Shannon entropy measures over the years and decades, with no reasonable explanation of how they are inserted into the hierarchy of fundamental measures led by Shannon entropy and algorithmic complexity, beyond simplistic and completely misleading accounts (such as Fig. 1 in~\cite{Marshall2022TheoreticalAT}).

\subsection*{The Assembly Index is no Better at Separating Peptides and Small Molecules}

\begin{figure}[htb]
    \centering
\includegraphics[width=0.9\textwidth]{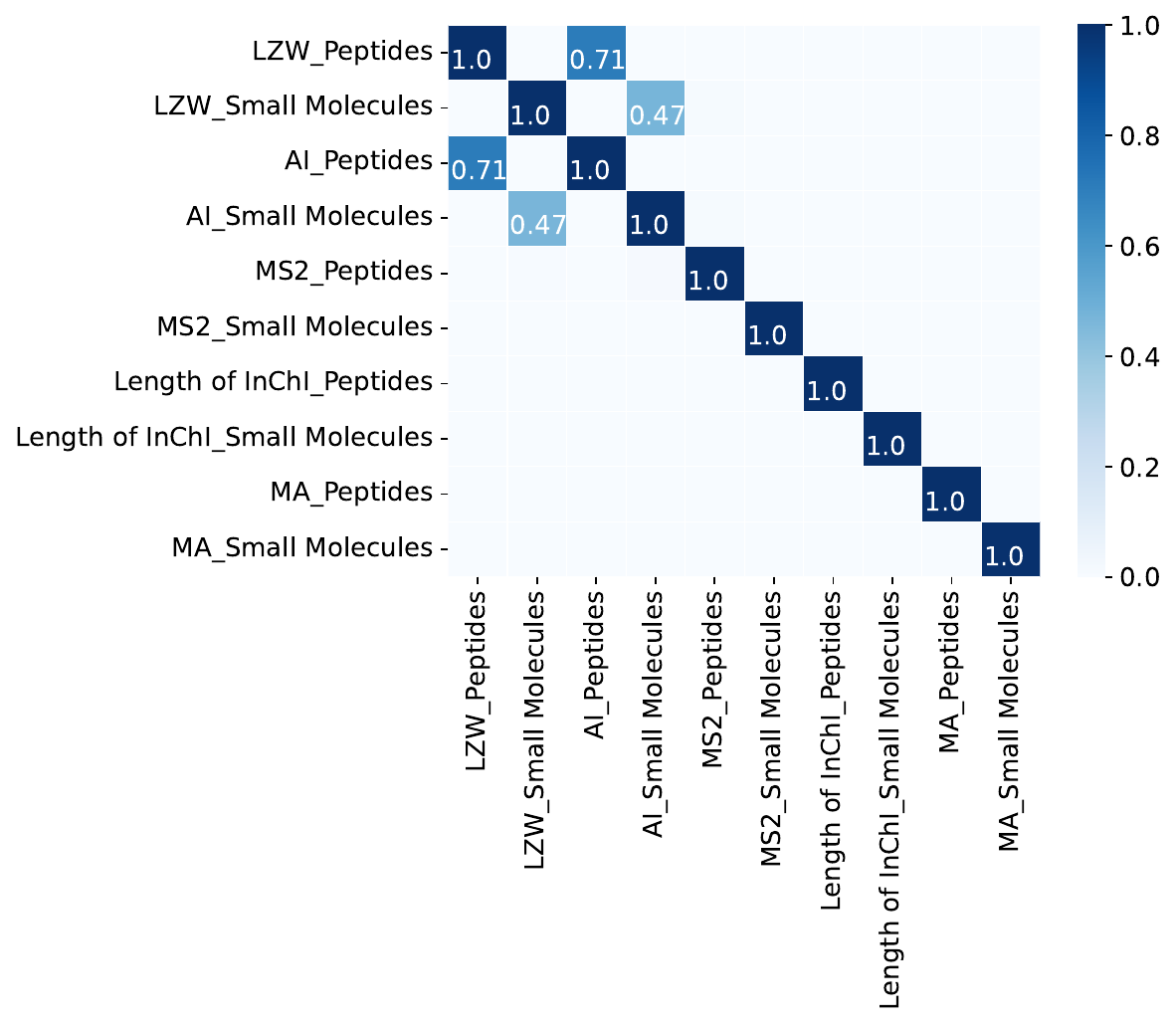}
    \caption{Heatmap showing the p-values for pair-wise Cucconi tests for all the groups presented in Fig.~\ref{fig:groupsMA}. The p-values greater than 0.05 were also explicitly written. It is clear that all the algorithms considered are able to separate between classes, because the pair-wise p-values reject the null hypothesis (equality of location and scale) with 95\% confidence. Also, in accordance with our previous claims, Ai and LZW (both for the same group of compounds) were tested for equal location and scale and such a null hypothesis was not rejected. This was conducted on the same small set of molecular compounds picked by the authors~\cite{Marshall2021} unlike the exhaustive experiment we reported years before~\cite{zenil2018}.}
    \label{fig:HeatgroupsMA}
\end{figure}

In order to compare the groups in Fig.~\ref{fig:groupsMA}, we perform a pairwise Cucconi test~\cite{cucconi,Marozzi-2009} to check if each pair can be considered to come from distributions with the same location and scale. The pair-wise p-values can be found in Fig.~\ref{fig:HeatgroupsMA}.

The p-values in Fig.~\ref{fig:HeatgroupsMA} can be used to assess if the groups formed, when analysed by the calculated metrics, are statistically similar. For example:

\begin{itemize}
    \item Let us take the pair ``LZW\_Peptides'' and ``LZW\_Small Molecules'' to compare.
    \item We may find the p-value for the comparison of these two groups in line 1, column 2 (or line 2, column 1). It is 0.0.
    \item For a 95\% level of confidence, we can say that the null hypothesis is rejected (p-value $<$ 0.05). Therefore, since the null hypothesis is that both groups have the same location and scale, we can say that the two groups are not statistically similar (location-scale wise, which is much stronger than simply comparing the means).
\end{itemize}

This analysis can be done for every other pair. In the case of ``LZW\_Peptides'' and ``AI\_Peptides'' (or ``LZW\_Small Molecules'' and ``AI\_Small Molecules''), the p-value is 0.71 (or 0.47), which implies that the null hypothesis is not rejected. Thus, this indicates that these pairs of groups can be considered statistically similar. 

It can be seen that all the algorithms used are able to form groups which reject the null hypothesis. Thus the naive algorithms utilised were able to properly separate the classes of Peptides and Small Molecules, and Ai exploited the same features that these naive statistical algorithms utilised without showing any advantage or special feature detection.

\begin{table}[h!]
\centering
\begin{tabular}{l|c|c|c|c}
\hline
\textbf{Algorithm} & \textbf{Shapiro\_P} & \textbf{Normality} & \textbf{Test\_Used} & \textbf{P\_Value} \\ \hline
MA                 & 5.85E-10            & Failed             & Kruskal-Wallis      & 2.92E-12          \\ \hline
LZW                & 1.57E-10            & Failed             & Kruskal-Wallis      & 1.26E-11          \\ \hline
Ai                 & 2.59E-10            & Failed             & Kruskal-Wallis      & 1.47E-11          \\ \hline
MS2                & 4.28E-05            & Failed             & Kruskal-Wallis      & 7.07E-12          \\ \hline
Length of InChI    & 4.07E-11            & Failed             & Kruskal-Wallis      & 8.58E-12          \\ \hline
\end{tabular}
\caption{Between group comparison results of various complexity measures from the box plot shown in~\ref{fig:groupsMA}. The normality tests (Shapiro-Wilk, given by P-values and a pass or fail for normality) were followed by the nonparametric Kruskal-Wallis test for comparing the complexity measures (algorithms) between independent samples: peptides and small molecules. All measures show significant differences between the two molecule types. Based on the P-values, MA can distinguish between the two categories of molecules with a performance comparable to the most naive statistical indexes and on all molecular descriptors (InChI length and MS2 spectra). This was conducted on the same small set of molecular compounds picked by the authors~\cite{Marshall2021}, unlike the exhaustive experiment we reported years before~\cite{zenil2018}.}
\label{statsboxplotbetween}
\end{table}

Statistically, as shown in Table~\ref{statsboxplotbetween}, there is no fundamental difference or advantage to using Ai over traditional methods, or utilising one or another data type, as we have shown that the main driver is simply molecular length and not any other feature, including `copy-number', not to mention all the false negatives that were misclassified by MA and were not highlighted as the ones above the mistaken and arbitrary threshold.

\subsection*{On Misleading Stochasticity Arguments in Assembly Theory}

Drawn from the assembly space (a directed multigraph), the assembly index suggests exploring and considering a multiplicity of pathways into its calculation.
This is effectively the same structure achieved by a nondeterministic CFG, or more generally a particular case of a nondeterministic Turing machine.
Notice that both cases are classical examples of computable processes in the same computation class as other universal computation models (in the Turing degree $ \mathbf{0} $).
In this regard, we highlight that:

\begin{enumerate}
\item The nature and number of alternative pathways does not play any role in the calculation of the assembly index once the shortest pathway is chosen. This means that the index is derived without weighing or taking into consideration any other pathway (as we do). 
This effectively makes it a version of the Minimum Description Length Principle as we have proven, because it is a rough approximation to algorithmic complexity (all the claims about causality and physicality amount to mechanical equivalence), but is implemented by a tool that is subsumed by Shannon entropy. Hence it is in turn encompassed by algorithmic probability or the universal probability (semi)measure (which are other indexes in algorithmic information theory) via the algorithmic coding theorem.
This is one of the fundamental results in AIT that connects stochastic processes to computable processes, and is overlooked by the authors of AT and~\cite{Wolpert2024StochasticProcessTuring}.

\item Assembly Theory would need to demonstrate that it can do something that other multiple stochastic measures cannot. Indeed, most complexity measures are stochastic in nature or by design. One argument could be that the assembly index combines this stochasticity with a mechanical assembly procedure, but it would still need to show it can do something that none of the other statistical measures, such as Shannon entropy, can do, or that it is also different from LZ algorithms. 
The evidence so far, however, is that when it comes to the tasks that Assembly Theory was supposed to innovate and pioneer, not only LZW but Shannon entropy and other naive statistical indexes are able to replicate or outperform AT.

\item AT would need to explain how and why selection and evolution would only entail a single, most simple operation, the join operation that AT is capable of capturing. Maybe regions of the genome known as `junk DNA' can partially be explained as the accumulation of concatenations of objects. But we already know that in evolutionary biology many other operations exist, such as transposition, reversion, complementation, deletion. Yet none of these operations can be quantified by AT. Only the joining of copies of simple things can be accounted for, something that cannot be justified theoretically or empirically and does not produce better results than might be obtained using equally naive algorithms.
\end{enumerate}

While~\cite{Wolpert2024StochasticProcessTuring} does not take sides, as is explicitly stated in the paper, we think it is attributing to AT some features that it does not have or does not fully exploit. While the authors of~\cite{Wolpert2024StochasticProcessTuring} do agree with us that AT can be (re)written in the context of AIT (as we have already done~\cite{abrahao2024,Uthamacumaran2024}), the authors of AT refuse to accept any connection to computer science (their words) as they claim mathematics and computer science are inadequate to describe the world, nature or any physical phenomena. It is one of our main criticisms that they have not only distanced themselves from the very thing they are, an approximation to algorithmic complexity with the limitation that all other statistical measures have, which is not a feature and is the property of assigning high randomness to low algorithmic complexity objects. This cannot be a feature because low algorithmic complexity objects are also the most causally packed, which is also what the authors of AT claim to quantify. In other words, they cannot quantify both things with such a measure and under these arguments.

Nevertheless, let us briefly entertain the possibility that AT is doing something different than, say LZ algorithms, because of its `stochastic nature'. It should then have an effect on its resulting calculations and produce results different from other nonstochastic approaches. However, this is what we have said all AT papers have lacked, and what is considered in science a basic control experiment, an elementary comparison to previous literature and existing knowledge, which AT has never supplied. However, we have, and the empirical results are that their measure is equivalent to LZ compression. When inspected formally, we also demonstrated that their definition of the assembly index is indeed an LZ grammar that approximates Shannon entropy at basically the same rate as LZ compression algorithms.

If AT takes into consideration or works in a different way yet correlates with the most popular compression algorithm to date and at the greatest statistical significance, what would the value of any feature attributed to it be? Even assuming such a stochastic feature, AT cannot capture anything beyond the realm of statistical triviality and therefore ends up performing just as any fully deterministic algorithm, as we have proven both empirically and theoretically. Thus it represents nothing fundamentally distinctive, nor does it provide any insights other than perhaps pedagogical or accidental insights. 
In other words, if AT is Shannon entropy implemented by an LZ compression algorithm with extra steps, why would researchers want the extra steps?

\begin{figure}[ht]
    \centering
    \includegraphics[width=1\textwidth]{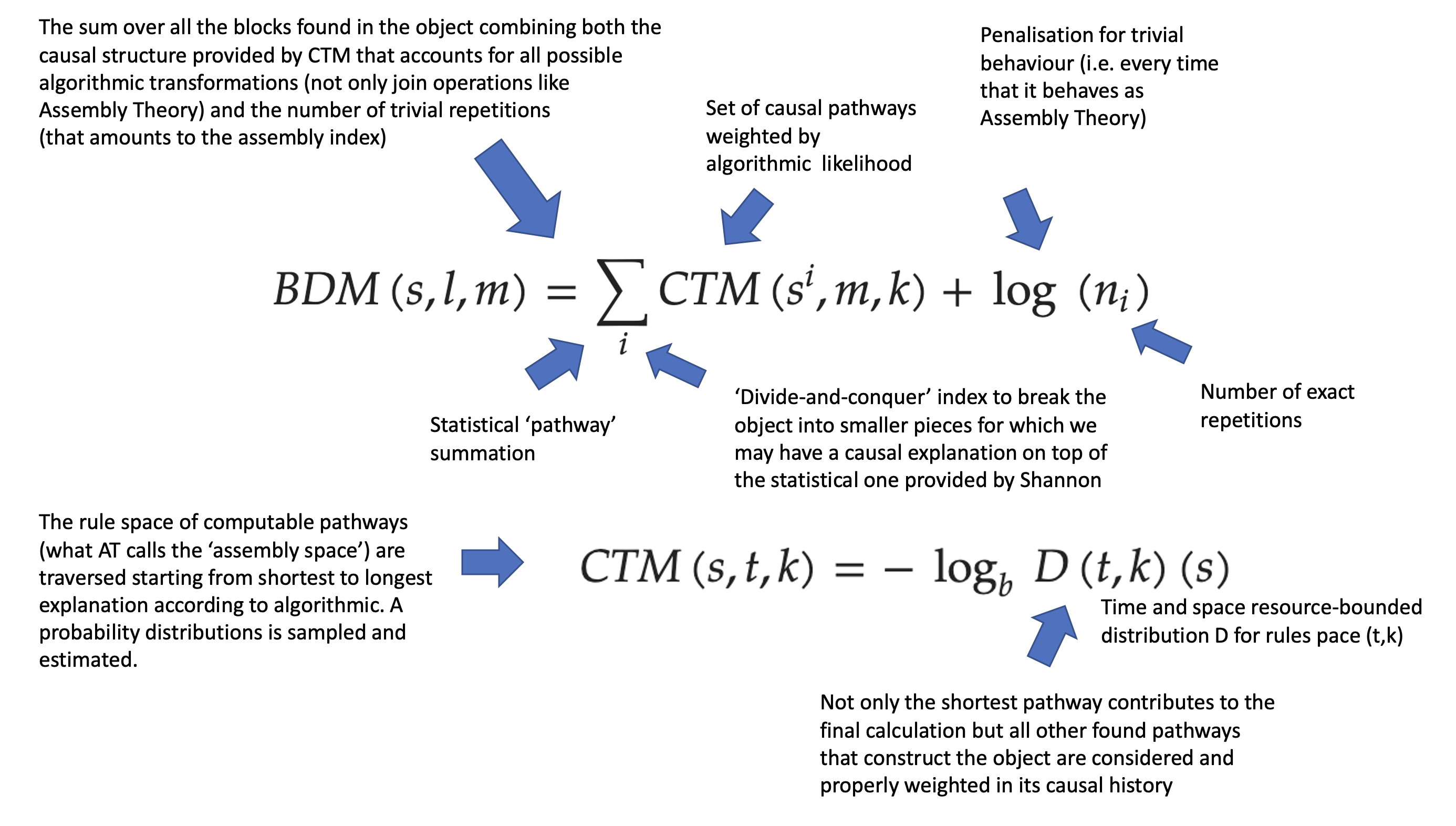}
    \caption{This diagram shows the similarities of our BDM that predates AT for many years and how its stochastic and computational components applied to any object $s$, finite, distinguishable and breakable (all properties that AT has deemed to innovate in dealing with when faced with having to define an object). 
    $s$ can be any type of object as we have shown before, not only binary strings but matrices, vectors, images, tensors, networks and objects in any vocabulary (binary or not). BDM traverses all the possible pathways from the most likely origin to the reconstruction of an object $s$ block by block, with the additional advantage that it can deal with and account for nontrivial causal regularities accounting for any operation and not only join operations as AT does, operations including but not restricted to deletion, subtraction, reversion, complementation, transposition, rotation, inversion, and so on. CTM, as an optimal inference method based on algorithmic probability, is the other side of the symbolic AI coin with BDM, the neurosymbolic approach that combines global statistical entropic distances with local symbolic algorithmic complexity distances. While there is a process to increase the accuracy of CTM at finding more and better computer programs, and such a process is semicomputable, CTM and BDM themselves are fully computable, and BDM runs in linear time.}
    \label{fig:bdm}
\end{figure}

The measures we introduced over a decade ago do what AT wanted to do but never did, and what others suggest that it does, but does not~\cite{Wolpert2024StochasticProcessTuring}. Our algorithmic measures~\cite{zenilctm,zenilbdm} are equipped to deal with both deterministic causal systems and the uncertainty and stochastic nature of the distribution of many possible explanations (or pathways). This is achieved by updating the belief system as regards which program is the best explanation for an observation once the observation differs from the output of the current program hypothesis, something we introduced in our Algorithmic Information Dynamics framework~\cite{iscience,nmi2019,aid1,aid2} based upon algorithmic probability, CTM~\cite{zenilctm} and BDM~\cite{zenilbdm}. This kind of Bayesian approach has been proven to be optimal and universal. At the same time, it is also computable, but not closed or trivial (as in only counting exact copies and accounting for a single join operation). It deals with model uncertainty and stochasticity according to the rules of current mathematical governing theories, namely classical information and algorithmic information. This is because we look at all the causal formal-theoretic explanations up to a certain size able to assemble an object by favouring those according to the universal probability distribution that encodes and formalises both the Principle of Multiple Explanations and the Parsimony Principle, which are the principles specifying not to ever discard any explanation but favour the simplest (shortest) ones.

In Fig.~\ref{fig:bdm}, the sum of all the pathways weighted by algorithmic probability (shortest and fastest first, longer later) on the basis that shorter explanations that can explain more of the object are favoured, which is consistent with the science dictum of model bias that can explain more with less and explored in our own papers(e.g.~\cite{bibid}). 
BDM can account for all causal operations and not only join operations like AT.
See also Section~\ref{sectionBDM}.

\subsection*{Uncomputability Phobia}

One regular argument made by the authors of Assembly Theory to defend their Assembly index against suggestions of similarities or full equivalence to algorithmic complexity is that their measure is computable, unlike algorithmic (Kolmogorov) complexity~\cite{pagel2024mapping,Marshall2022TheoreticalAT,kempes2024}. They have even provided a several-page mathematical proof to this effect. Clearly, the index did not need any proof because it is trivially computable, and this is now even more evident given their fully proven mathematical equivalence to popular lossless compression LZ algorithms that are upper bounded by Shannon entropy, which is trivially computable as well.

On the one hand, algorithmic (Kolmogorov) complexity is lower semi-computable, meaning it can be approximated from below, and some compression is sufficient proof of nonrandomness. On the other hand, algorithmic complexity is semicomputable if the only answer accepted is the smallest computer program, but neither applications of algorithmic complexity nor the Assembly index are exclusively concerned with the shortest possible algorithm (or pathway). This means that finding any shorter pathway or computer program is computable. Moreover, for decades, researchers have introduced resource-bounded complexity measures based on or inspired by the same arguments of conciseness of mechanistic description, including our own Coding Theorem and Block Decomposition Methods~\cite{zenilctm,zenilbdm}, which are fully computable, but also LZ(W) algorithms used in all sorts of applications for characterisation and identification of chemical and physical properties of life, selection and evolution~\cite{cilibrasi,Li1997}.

Many of these misguided arguments~\cite{Marshall2022TheoreticalAT}, however, seem to arise from a phobic reaction to uncomputability (or semi-computability) processes or methods. However, we also have introduced a computable but stochastic method in AIT that is compatible both with classical information theory and algorithmic complexity based on the principles of algorithmic probability. Therefore, AT offers no advantage, and it has been proven to be subsumed within AIT~\cite{abrahao2024}.
Thus, we agree with the authors in~\cite{Wolpert2024StochasticProcessTuring} that AT can be fully framed in the language of AIT, with or without the stochasticity argument. But this is exactly one of our main criticisms of AT, and hence we fully agree with the criticism made in~\cite{Wolpert2024StochasticProcessTuring} against AT, which claimed this theory would have nothing to do with computer science, compression or `Turing machines' because, according to the authors of AT, these fields cannot contribute or deal with `physical data' or be relevant to `experiments'.

While it is difficult to understand what this may mean, it clearly was not the opinion of the Nobel committee when awarding the 2024 Nobel Prize in Physics and Biology to AI and computer science groups for their contributions to physics and biology. Moreover, the
spectacular effectiveness 
of mathematics and computer science 
in comprehending the real-world~\cite{bibid} is widely known,
and we could cite thousands of papers of applications from computer science to the physical sciences (e.g.~\cite{ren1995construction,chaitankar2010novel}) in highly experimental settings. 
 Without computer science algorithms, many fundamental tasks in the experimental sciences would be impossible. For instance, we rely on algorithms to efficiently sort and analyse experimental data, filter and process signals in various fields like neuroscience or seismology, reconstruct images in MRI scans for medical diagnostics, and capture and analyse spectral data to study distant stars and galaxies in astrophysics. Furthermore, the Assembly index is itself an algorithm written in a basic language of mathematics, implemented as a simple computer program, and run on a digital computer reading a computer file. The only plausible explanation seems to be a flawed assumption that since the authors of Assembly Theory (AT) are physicists and chemists, only their algorithms can address phenomena related to physics or chemistry (and, by extension, biology, selection, evolution, and more).

In contrast, the measure we introduced over a decade ago~\cite{zenilctm,zenilbdm} combines the current state-of-the-art algorithms from which it draws its power or is based upon, does take into consideration all the possible pathways along which an object could have been constructed causally and mechanically, and gives proper weight to the many other pathways (not only the shortest one as in AT), thereby even implementing what the AT authors often claim to have implemented but did not. What they have done is aim directly for a statistical approximation of algorithmic complexity, with the known limitations of such approximations, which they present as features, limitations that all statistical Shannonentropic indexes possess. All the while they've claimed to be different, even when such a claim is mathematically disproven ~\cite{Marshall2022TheoreticalAT}. Indeed, they have not yet proven that they can do anything that Shannon entropy, Minimum Description Length (MDL) approaches or LZW and cognates can already do and have done for the last 50 years or more (e.g.~\cite{saitou}).

Were we to give in to 
this obsession to avoid uncomputability in some academic circles (e.g.~\cite{Marshall2022TheoreticalAT}), mostly a product of naivety or ignorance, we would have to
abandon entire areas of modern mathematics and physics, 
from calculus to set theory to molecular dynamics to category theory to mathematical analysis to quantum logic in favour of, say, Euclidian geometry, Presburger arithmetic, or Greek atomism.
These may not be uncomputable but they are also so restricted in what they can represent and explain that they are utterly inadequate when confronted with the full scope of the phenomena falling within the purview of geometry, mathematics and physics. 
Indeed, the greatest expressibility is achieved through computational universality, which results from putting very simple operations together, like simple arithmetic, leading to uncomputability through universality. This uncomputability should be a desired feature rather than a negative or undesirable quality. 

This fear of uncomputability is what Assembly Theory evinces, indulging it by adopting an index that can only account for concatenation and join operations, though it cannot explain 
even the most basic of the arithmetical transformations or implement a basic logic Boolean circuit, sending us back 50 to 70 years in time to the beginnings of Shannon entropy and complexity theory.

This uncomputability phobia stems from a combination of ignorance and conformism. There is a lack of understanding of the fundamental role that uncomputable problems and uncomputable methods play in areas of science, helping to shape and advance solutions in both theoretical and experimental sciences.

This often arises from an academic culture that tends to prioritise the safe, well-defined boundaries of redundant methods for trivial problems over stronger methodological foundations. If we are to have any chance of dealing with causality, for example, there is no option but to deal with the uncomputability that arises from asking nontrivial questions about simple systems that are capable of universal computation~\cite{bibid,nmi2019}.  The fear of uncomputability reflects a narrowed scientific understanding of open-ended methods equipped to deal with open-ended challenges, such as the causal challenge of inverse problems.

As of today, the authors keep repeating falsely~\cite{seet2024rapid} that these measures related to algorithmic (Kolmogorov) complexity are uncomputable (even citing measures that are known to be computable from Cilibrasi, Vitanyi and others) just to wrongly claim the opposite, and to avoid any reference to the whole literature on resource-bounded algorithmic complexity, spearheaded
by our group, with applications to exactly the same domains as Assembly Theory (sans the hyperbolic claims), while they also partially backtrack on some of their previously incomprehensible statements claiming the Assembly index to be unrelated to computer science, compression, or even any algorithms. What this amounts to is an unapologetic rewriting of their own history as a possible positive result of our recurrent criticisms and proofs against their multiple unfounded claims~\cite{Abrahao2017publishednat,Uthamacumaran2024}. 


Finally, they also often use the argument that most computer science deals only with toy examples relevant only to computer science itself~\cite{Marshall2022TheoreticalAT}, such as Turing machines. However, the concept of a Turing machine arose from a completely opposite effort: to find the most general of possible objects by proving universal equivalence. 

Indeed, a universal Turing machine proves that any algorithm can be written as a Turing machine and vice versa, and an algorithm is the colloquial term we use in science for any method or rule-based procedure. It can hardly be more general or about the most general of the concepts of scientific application. The Assembly index is only one such algorithm that can be described as a Turing machine, and a very weak one indeed, as being a finite state automaton, it does not need to account for any operation but concatenation, disregarding everything else. Computer science is, therefore, not an area suitable only for computer science but for science in general, and the Swedish Nobel Academy recently awarded its Chemistry and Physics prizes for the contributions of computer science to these sciences.

\subsection*{Clustering Molecules from Experimental Data}

In an exercise to investigate the extent to which the results that Ai suggests are unique to their theory, which, they claim, supports their view that Assembly Theory can unify physics and biology, explain selection and evolution and characterise life on Earth and beyond~\cite{Marshall2021,Sharma2023}, we took the data from~\cite{Marshall2021} and reproduced their analysis in Fig.~\ref{fig:AllQuantiles} with statistical measures, including some naive and the most basic Machine Learning and Deep Learning approaches, supplying the basic control experiments for comparative purposes that the original authors neglected to provide.

\begin{figure}[htbp]
    \centering
\includegraphics[width=0.8\textwidth]{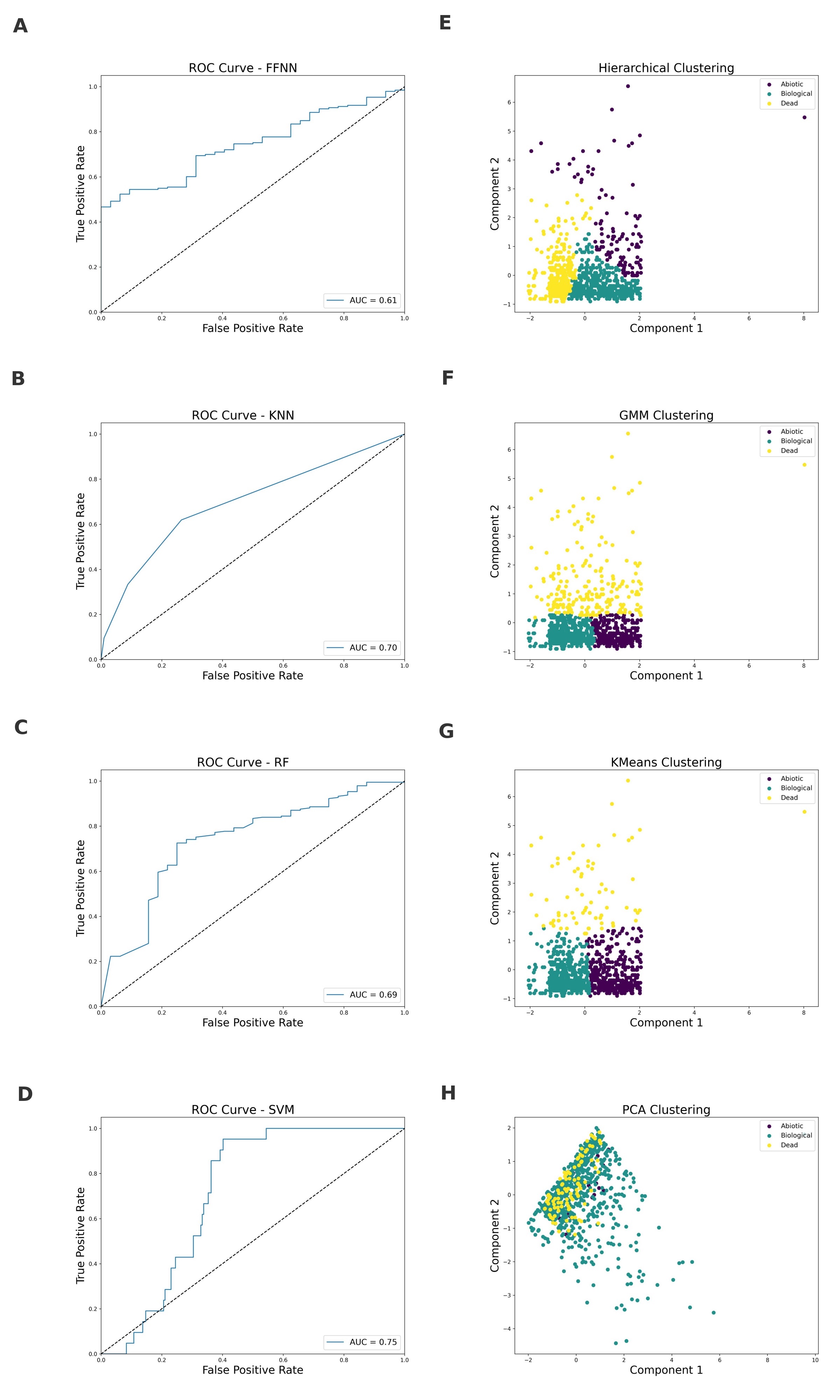}
    \caption{Clustering and machine learning classification performance on mass spectrometry peak matrices from source data showing that any naive statistical or machine learning algorithm (for both transparent- and opaque-box approaches) can reproduce the results that Assembly Theory deems the ones unifying biology and physics~\cite{Sharma2023}.
    }
    \label{fig:ml_measures}
\end{figure}

Fig.~\ref{fig:ml_measures} on the right-hand side, shows four different clustering algorithms to visualise how well the mass spectra peak matrices (i.e., m/z ratio vs. a number of peaks) cluster by the three categories. As shown, hierarchical clustering, Gaussian Mixture Models (GMM), KMeans cluster, SVM, Random Forest (RF), KNN, and a neural network (FFNN) or multilayer perceptron, all but PCA 
separate the three categories.

    In Fig.~\ref{fig:ml_measures}, from A to D on the left-hand side, we see the ROC curves of four ML algorithms classifying the organic and inorganic molecules (N=30) from the Marshall et al.~\cite{Marshall2021} paper's Fig. 4 mass spectrometry mixtures into 3 categories: Dead, Inorganic, and Biological. SVM had the highest area under the curve (AUC), with 0.75, and a classification accuracy of 0.86.
    SVM suggests linear hyperplane separability. 

\begin{figure}[htbp]
    \centering
    \includegraphics[width=1.0\textwidth]{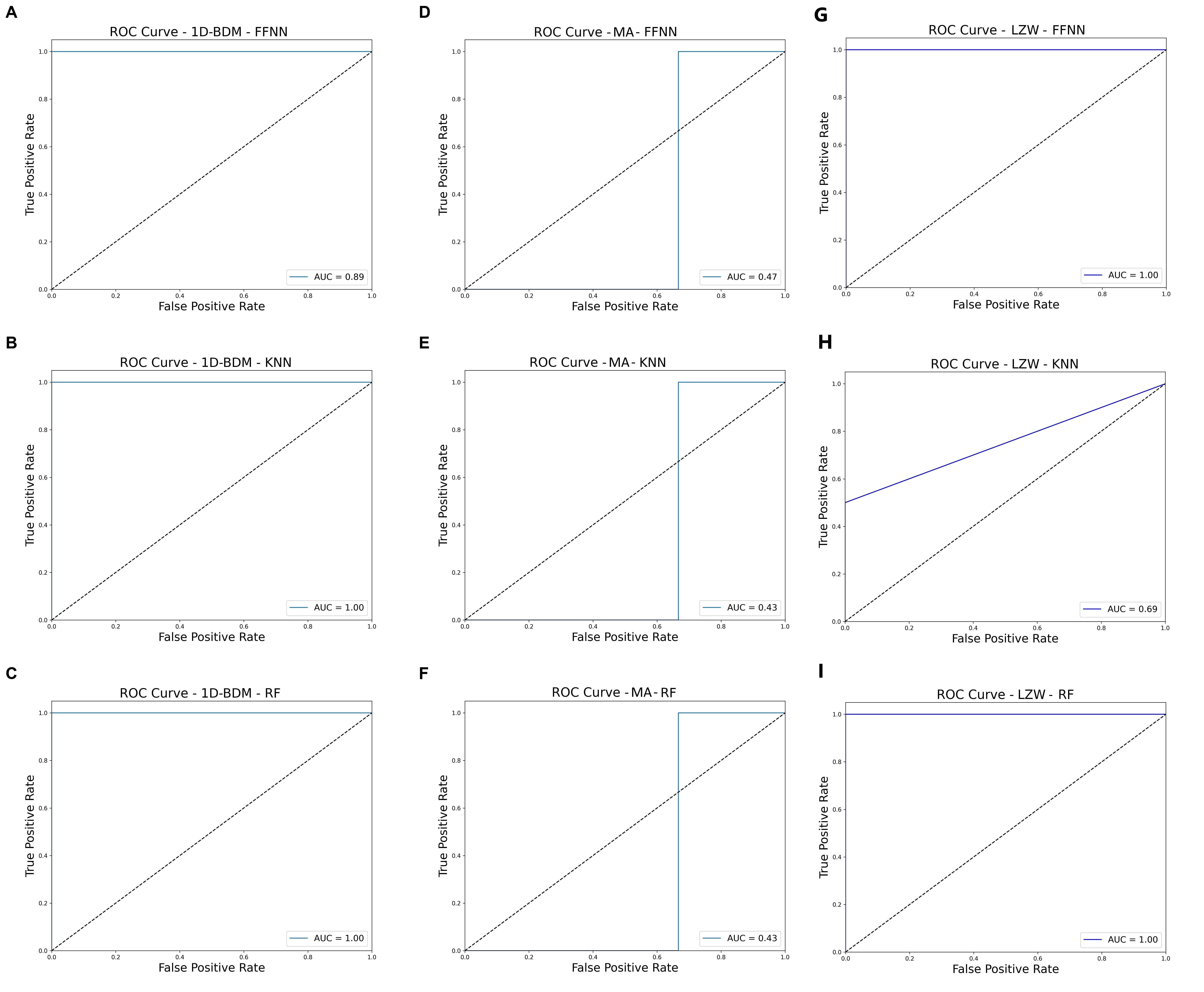}
    \caption{ROC curves comparing mass spectrometry data of organic and inorganic molecules by way of comparing the measures of 1D-BDM and MA. LZW, and other compression schemes like RLE and Huffman, had similar values close to the evaluated ML performance of BDM (AUC scores of 0.89 for MLP and 1.00 for KNN and RF). It is clearly shown that BDM (and, similarly, other AIT measures) outperforms MA (the values attributed to the environmental samples' mass spectrometry data in Marshall et al.~\cite{Marshall2021} Figure 4) using basic machine learning algorithms. Likewise, LZW compression also outperforms MA, with a perfect AUC (1.00) using FFNN and RF and a better AUC than MA for KNN (0.69). MA has poor AUC scores of 0.47 for MLP and 0.43 in KNN and RF-based classification. All this even gives AT and MA the advantage of using their own small choice of molecules~\cite{Marshall2021} rather than the exhaustive experiment we reported years before~\cite{zenil2018} on more than 15\,000 compounds.}
    \label{fig:roc_curves}
\end{figure}

    The ML predictions are further supported by the ROC curves in Figure~\ref{fig:roc_curves}. Fig.~\ref{fig:roc_curves} demonstrates that 1D-BDM and LZW compression measures significantly outperform  MA as complexity measures in distinguishing between inorganic, dead, and biological molecules from~\cite{Marshall2021} mass spectrometry data. Specifically, the AUC scores for BDM and LZW approaches are perfect or near-perfect (AUC = 1.00), while MA struggles with considerably lower AUC scores: within the 0.43-0.77 range with ML classification. This underperformance suggests that MA is suboptimal for detecting and quantifying biosignatures, possibly due to its less robust representation of molecular complexity in comparison to complexity measures from the AIT framework, like BDM and LZW, which more effectively capture the intricate patterns needed to differentiate between these categories. Consequently, MA (Ai) may not be well-suited for the study's objective of identifying unique biosignatures of life based on molecular complexity and its multi-scale organisation.

Therefore, these findings demonstrate that even naive ML approaches could achieve equal or superior classification performance compared to the assembly index (Ai), with SVM and FFNN achieving high precision and AUC scores, thus challenging the uniqueness and utility of Ai as a biosignature discriminator.




\end{appendices}
\end{document}